\pdfoutput=1
\documentclass[useAMS,usenatbib]{mn2e}
\usepackage{amssymb}
\makeatletter
\newcommand*{\rom}[1]{\expandafter\@slowromancap\romannumeral #1@}
\makeatother
\usepackage{tabularx}
\usepackage{graphicx}
\usepackage{ulem}
\usepackage{array}
\usepackage{booktabs}
\usepackage[fleqn]{amsmath}
\usepackage{subfigure}

\title[Disc heating by satellite galaxies]{Impact of cosmological satellites on the vertical heating of the Milky Way disc}

\author[R. Moetazedian $\&$ A. Just]{R. Moetazedian\thanks{E-mail:
reza@ari.uni-heidelberg.de (RM); just@ari.uni-heidelberg.de (AJ)}\thanks{Fellow of the International Max Planck Research School for Astronomy
and Cosmic Physics at the University of Heidelberg (IMPRS-HD).} and A. Just\footnotemark[1]\\
Astronomisches Rechen-Institut, Zentrum f\"{u}r Astronomie, University of Heidelberg, M\"{o}nchhofstr. 12-14, 69120 Heidelberg, Germany\\}

\begin{document}

\date{Accepted 2016 March 31. Received 2016 March 24; in original form 2015 August 14}

\pagerange{\pageref{firstpage}--\pageref{lastpage}} \pubyear{2015}

\maketitle

\label{firstpage}

\begin{abstract}
We present a high resolution study of the impact of realistic satellite galaxies, extracted from cosmological simulations of Milky Way haloes including 6 Aquarius suites and Via Lactea \rom{2}, on the dynamics of the galactic disc. The initial conditions for the multi-component Milky Way galaxy were generated using the GalIC code, to ensure a system in dynamical equilibrium state prior to addition of satellites. Candidate subhaloes that came closer than 25\,kpc to the centre of the host DM haloes with initial mass enclosed within the tidal radius, $M_\textrm{tid}$ $\ge$ 10$^{8} M_{\odot}$\,=\,0.003 $M_\textrm{disc}$, were identified, inserted into our high resolution N-body simulations and evolved for 2 Gyr. We quantified the vertical heating due to such impacts by measuring the disc thickness and squared vertical velocity dispersion $\sigma_{z}^{2}$ across the disc. According to our analysis the strength of heating is strongly dependent on the high mass end of the subhalo distribution from cosmological simulations. The mean increase of the vertical dispersion is $\sim$ 20\,km$^{2}$\,s$^{-2}$\,Gyr$^{-1}$ for R $>$ 4\,kpc with a flat radial profile while, excluding Aq-F2 results, the mean heating is $<$ 12\,km$^{2}$\,s$^{-2}$\,Gyr$^{-1}$, corresponding to 28\% and 17\% of the observed vertical heating rate in the solar neighbourhood. Taking into account the statistical dispersion around the mean we miss the observed heating rate by more than 3$\sigma$. We observed a general flaring of the disc height in the case of all 7 simulations in the outer disc.
  
\end{abstract}


\section{Introduction}
\label{sec:int}
According to the $\Lambda$ cold dark matter ($\Lambda$CDM) paradigm, structure formation follows a hierarchical scenario; dark matter halo growth occurs via the accretion of smaller size systems (e.g. \citealt{white-rees,moore}).
High-resolution cosmological simulations have demonstrated that DM substructures survive within environments similar to the Milky Way\,\citep{moore}. Galaxy mergers are believed to be the main drivers for significant kinematical 
perturbations of galaxies\,\citep{toomre}; these can be classified into three categories. With regards to theoretical\,\citep{hernquist1} and observational evidences, (e.g. \citealt*{woods1,woods2}) ``major'' mergers (mass ratios $\gtrsim$ 1:3 of total galaxy mass)  strongly perturb the system and can even change the morphology. However such events are less common compared to ``minor'' mergers with masses in the range 1:3 - 1:50 of total mass -- 1:1 of disc mass -- which are likely to destroy thin discs\,\citep{kazan08}. The third type, which forms the basis of this study, is the interaction of satellite galaxies with intermediate mass ratios which are more commonly represented in units of the disc mass ($\lesssim$ 1:10 of disc mass). Our Milky Way galaxy has been experiencing such mergers; observations of stellar tidal stream remnants  provide evidence for such encounters (e.g. \citealt*{ibata1,majewski,martinez}). Understanding the evolution of disc galaxies has challenged both observers and theorists for the past two decades. There have been numerous attempts at bridging the observed characteristics of spiral galaxies and their violent history (e.g. \citealt{navarro-white,abadi}). Some of the early studies, including that by\,\citet{moore} indicated that interaction of substructures with galactic discs could induce heating processes resulting in thickening of the disc. Also the $\Lambda$CDM scenario predicts a continuous infall of subhaloes -- including massive satellites, which are expected to impact the dynamics of the stellar disc significantly\,\citep{stewart}. However early observations of our solar neighbourhood  found the presence of a 10 Gyr old ``thin'' disc with a scale height $<$ 500\,pc (e.g. \citealt{kent,dehnen,just}).

The survival of a thin disc in the ($\Lambda$CDM)  universe has been tested through early analytic models (e.g. \citealt{toth,sellwood}) together with N-body/hydrodynamical simulations (e.g \citealt*{quinn2,quinn,walker,velazquez,sommer}). One of the major shortcomings of analytical approaches was the choice of nearly circular satellite orbits. This is not in agreement with the $\Lambda$CDM results, where eccentric orbits are favoured\,\citep{ghigna}. Another disadvantage was the use of a rigid potential for the host halo, disc and/or bulge instead of live components. Rigid potentials suppress the effect of interactions between the halo and the disc, as well as the dynamical friction --  and hence transfer of angular momentum -- between different components as the satellites spiral towards the centre.\,\citet{velazquez} have shown that using  a rigid potential instead of a live halo overestimates the heating of the galactic disc by a factor of 1.5-2. Early numerical simulations suffered from an  unrealistic high abundance of substructure compared to the $\Lambda$CDM cosmology prediction. In addition to  galaxy mergers, other proposed  mechanisms are expected to contribute towards the heating. For example,\,\citet{carlberg87} and\,\citet{sellwood13} looked into the induced changes of disc's phase-space distribution due to transient spiral waves and concluded that such structures are able to increase the velocity dispersion of stars, although in the disc plane, an increase of the velocity anisotropy and hence smaller $\sigma_{z}/\sigma_{R}$ is observed.  Scattering due to Giant Molecular Clouds (GMCs) is another possible scenario\,\citep{kokubo}. An overview of possible disc heating mechanisms are discussed in the review by\,\citet{sellwood14}.

More recent simulations (e.g. \citealt{font,ardi,kazan08,read,villalobos,kazan09}) have shown that infall of satellite galaxies could result in either destroying or greatly heating the disc, forming a \textit{thick} component; this contradicts our picture of the Milky Way thin disc component. It is important to note these analyses focussed on the impact of satellites at the massive end of the substructure spectrum, with masses $>$ 5 $\times$ 10$^{9}$ $M_{\odot}$. For instance,\,\citet{read} concluded that accretion of a satellite with mass comparable to the Large Magellanic Cloud (LMC) -- 10$^{10}$ $M_{\odot}$ -- is sufficient to explain the thick disc formation. Also\,\citet{moster} examined the role of a gas component in disc instability and found, that for a gas mass fraction of 25 percent, the thickening of the disc could decrease by 20 percent -- thus suppressing vertical heating. In addition,\,\citet{quillen} and\,\citet{bird} showed that satellites infall could also induce radial migration in MW-like systems with greater impact at larger radii, while\,\citet{minchev12} and\,\citet{vera-ciro} concluded that radial migration has minimal impact on the thickening of the MW disc. The radial migration as a result of mergers reduces the disc flaring\,\citep{minchev14} which provides an opposite picture to the measured one, especially in the outer parts, in the absence of external perturbations as demonstrated by\,\citet{minchev12}. Based on perturbation theory,\,\citet{vesp00} have shown that energy deposit in the perturbation by satellites, scales as the square of the satellite mass, and that in the resonance situation, minor mergers may have a measurable impact on the host galaxy. In a cosmologically based hydrodynamical simulation, \citet{gomez} have shown that even a distant encounter of a satellite galaxy with $M\,=\,4\times10^{10} M_{\odot}$ ($\sim$ 5 per cent of the host galaxy mass) at R\,=\,80\,kpc can perturb the outer disc at R $>$ 12\,kpc significantly and form a Monoceros ring-like warp. 

The observations of the local stellar kinematics have suggested multiple scenarios for the evolution of the thin disc. The most useful observational quantity in order to investigate the evolution of our Milky Way disc is the so-called ''age-velocity dispersion'' relation. This relation is aimed at connecting stars' kinematics with their dynamical history and providing constraints on the vertical structure of the Galactic disc.\,\citet{nord} and follow-up studies (e.g. \citealt*{holmberg07,holmberg09}) have favoured a picture in which the disc has experienced  continuous heating through its lifetime. Another possible scenario is the saturation of vertical heating of stars older than 4.5 Gyr, as suggested by a detailed analysis of the Geneva-Copenhagen survey\,\citep{seabroke}.\,\citet{holmberg09} fitted power laws to the radial, tangential, vertical and total velocity dispersions of $\thicksim$ 2600 F and G dwarfs in the solar neighbourhood and derived power law exponents of $\gamma$\,=\,0.39, 0.40, 0.53 and 0.40 for these respective quantities.    

Within the framework of this study, we use high-resolution N-body simulations\,\citep{bien} to investigate the impact of accreting satellite galaxies on the dynamics of our Milky Way ``thin'' disc and answer the question: are we able to explain the observed heating of the Galactic disc in the presence of satellites infall? For improved realism, the initial conditions (ICs) of our satellites follow that of the $\Lambda$CDM model\,\citep{aquarius}. The interactions of satellites and the stellar disc have the potential to trigger different types of both small and global scale perturbations, such as initiating modes in the Galactic disc. Hence, for a deeper understanding of such distortions in the phase-space distribution of disc stars, we require simulations that are capable of resolving the dynamics down to few tens of \textit{parsecs} scale; a resolution  much higher than most current Milky Way interaction analysis simulations. We employ multi-component models for our Milky Way which enable us to use both observational and cosmological constraints for modelling our Galaxy\,\citep{yurin}.

This work improves the current knowledge about the impact of subhalo interactions on the observed vertical structure of our Galactic disc. Such structure is the result of a complex interplay of different heating mechanisms and, since we do not yet know about the exact contribution from each of such mechanisms,  quantifying one will provide a more accurate picture of other processes. For the first time, we have analysed and compared the impact of subhaloes from Aquarius and Via Lactea \rom{2} Milky Way-like simulation suites (e.g. \citealt{aquarius,diemand1})  to provide a statistically valid study of satellite impact on the vertical structure of the stellar disc of our Galaxy. In addition, having initial conditions for the MW multi-component galaxy which are in much better equilibrium state, compared to previous work, allows us to minimize the bias effect which, in principle, could overshadow the expected real heating measurements. The combination of 40\,pc vertical resolution together with ICs in a robust equilibrium state differentiates this analysis  compared to  previous studies -- allowing  minor vertical heating effects to be measured and distinguished from other processes.

In section\,\ref{sec:sim-suite} we discuss the cosmological simulation suites that we use for the purpose of our realistic initial conditions of DM structures together with the statistics of such objects. The N-body model used for the multi-component Milky Way and satellites, a description of our N-body code and the isolated systems simulation results are explained in section\,\ref{sec:model}. The results of simulations with satellites for the solar neighbourhood and across the disc, together with an analysis of the orientation of subhalo distribution, are discussed in section\,\ref{sec:results} and we finish the paper with a summary in section\,\ref{sec:summary}.

\section[]{Cosmological Simulation Suites}
\label{sec:sim-suite}
For a realistic picture of the interaction between satellite galaxies and the Milky Way's thin disc we have extracted the distribution of subhaloes from two suites of cosmological simulations: Aquarius and Via Lactea \rom{2} (e.g. \citealt{aquarius,diemand1}). The major difference between the Aquarius and Via Lactea \rom{2} (hereafter VLII) lies in  their adopted cosmology, which in the case of Aquarius are the WMAP one-year results\,\citep{klypin}, while for VLII WMAP three-year results were used. Hence the value of $\sigma_\textrm{8}$, which represents the amplitude of the (linear) power spectrum on the scale of 8 h$^{-1}$Mpc, is lower for VLII. This parameter plays a crucial role in the context of influencing structure growth at early epoch. Also n$_\textrm{s}$ corresponds to the spectral index of scalar fluctuations (Table\,\ref{tab:1}) and is close to being scale-invariant. The value for $\Omega_\textrm{m}$, $\sigma_{8}$ and n$_\textrm{s}$ are smaller for VL\rom{2} compared to Aquarius which might impact the substructure distribution at the very low mass end. 
\subsection{Host dark matter halo properties}
\label{sec:dm-halo}
\begin{table}
 \caption{the cosmological initial conditions} \label{tab:1}
\begin{tabular}{cccccc}
\hline
Cosmological Suite & $\Omega_\textrm{m}$ & $\Omega_\textrm{$\Lambda$}$ & $\sigma_\textrm{8}$ & n$_\textrm{s}$ \\ \hline
Aquarius & 0.25 & 0.75 & 0.9 & 1.0 \\ 
VL\rom{2} & 0.238 & 0.762 & 0.74 & 0.951 \\ \hline
\end{tabular}
\end{table} 
The Aquarius suite consists of a set of 6 simulations; each associated with a different realisation of DM halo likely to host a Milky Way-like galaxy and which has not experienced a recent major merger. The haloes are chosen from a parent halo of homogeneous resolution. These simulations are performed in cubic box with side length of 137 Mpc while VL\rom{2} is performed in a smaller box of length 40 Mpc. The second highest resolution set of Aquarius simulations ``Aq-A2$\cdots$Aq-F2'' are employed for this analysis. In our study we use the \textit{z}\,=\,0 snapshots from all the above simulations, which resemble the present day distribution of subhaloes. Each snapshot contains information for the subhaloes -- including the position, velocity, maximum circular velocity V$_\textrm{max}$, position of this velocity $r_\textrm{v$\textrm{max}$}$ and tidal mass $M_\textrm{tid}$ for every substructure residing within the simulation box. 

The parent haloes from these seven cosmological simulations do not possess exactly similar profile characteristics such as $M_\textrm{200}$, the mass enclosed within the sphere with radius $r_\textrm{200}$, where $r_\textrm{200}$ represents the radius at which the mean density of the DM halo is 200 times the critical density of the universe ($\rho_\textrm{crit}$). Thus, we  scale the properties of subhaloes from all seven simulations in terms of their position, velocity and mass using similar scaling relations mentioned by\,\citet{kannan}. All the simulations were scaled with respect to the $M_\textrm{200}$ of the ``Aq-D2'' simulation - 1.77 $\times$ 10$^{12}$ $M_{\odot}$. The quantity \textit{g} represents the scaling factor in
\begin{equation}
  M=M_\textrm{orig}/g,
  \label{eq:scl1}
\end{equation} 
\begin{equation}
  v_\textrm{\textit{x,y,z}}=\frac{v_{\textrm{orig},x,y,z}}{g^{1/3}} \qquad \textrm{and}
  \label{eq:scl2}
\end{equation} 
\begin{equation}
  x_=\frac{x_\textrm{orig}}{g^{1/3}}
\qquad
  y=\frac{y_\textrm{orig}}{g^{1/3}}
\qquad
  z=\frac{z_\textrm{orig}}{g^{1/3}}.
  \label{eq:scl3}
\end{equation} 
The subscript ``orig'' corresponds to the original non-scaled quantities. Such scaling is important for a statistically meaningful comparison of different simulations. After having done the rescaling, all the host haloes have the same mass and size but different concentrations, hence different $r_\textrm{vmax}$ and $V_\textrm{max}$. 

According to early N-body simulations of dark matter haloes following the $\Lambda$CDM structure formation model, the density profile of the DM haloes could be well described via the known Navarro-Frank-White (NFW) profile\,\citep{nfw} 
\begin{equation}
  \rho_\textrm{NFW}(r)=\frac{\rho_\textrm{s}}{(r/r_\textrm{s})(1+r/r_\textrm{s})^2}.
  \label{eq:nfw}
\end{equation}
where $\rho_\textrm{s}$ and $r_\textrm{s}$ are the scale density and scale radius of the subhalo. The $\rho_\textrm{s}$ can be computed given the concentration c of the halo which is simply the ratio of $r_\textrm{200}/r_\textrm{s}$ and the overdensity $\delta_\textrm{c}$ with respect to $\rho_\textrm{crit}$ using 
\begin{equation}
 \delta_\textrm{c}=\frac{\rho_{s}}{\rho_\textrm{crit}} = \frac{200}{3}\frac{\textrm{c}^{3}}{ln(1+\textrm{c})-\textrm{c}/(1+\textrm{c})}
  \label{eq:deltac}
\end{equation}
%
%
Table\,\ref{tab:2} shows the \textit{normalised} properties of the host haloes from VL\rom{2} and six Aquarius simulations. The particle resolution of the simulation suites m$_\textrm{p,orig}$ which is $\sim$ 10$^{4}$ $M_{\odot}$ motivated us to perform a mass cut and consider subhaloes with $M_\textrm{tid}$ $\ge$ 10$^{6}$ $M_{\odot}$ since substructures with lower mass consist of only a few hundred particles. This introduces some uncertainties in determining the characteristics of these subhaloes at the outer regions. The  analysis performed on the subhaloes in this paper include this lower mass cut. In addition to $M_\textrm{200}$ and $r_\textrm{200}$, the maximum circular velocity V$_\textrm{max}$ and concentration c are listed. For each parent halo we also know $r_\textrm{50}$,  the radius at which the mean density is 50 times $\rho_\textrm{crit}$; V$_\textrm{50}$, the circular velocity at this radii and $f_\textrm{sub}=f_\textrm{sub}^{cum}(r_{50})$, the percentage fraction of the cumulative mass residing in substructures relative to the host halo enclosed mass at $r_\textrm{50}$ and seen as a measure of the host halo clumpiness. It appears that VL\rom{2} has less mass residing in bound substructures -- $f_\textrm{sub}$ $\thicksim$ 5 percent -- compared to the Aquarius simulations. 

\begin{figure}
\includegraphics[width=\linewidth]{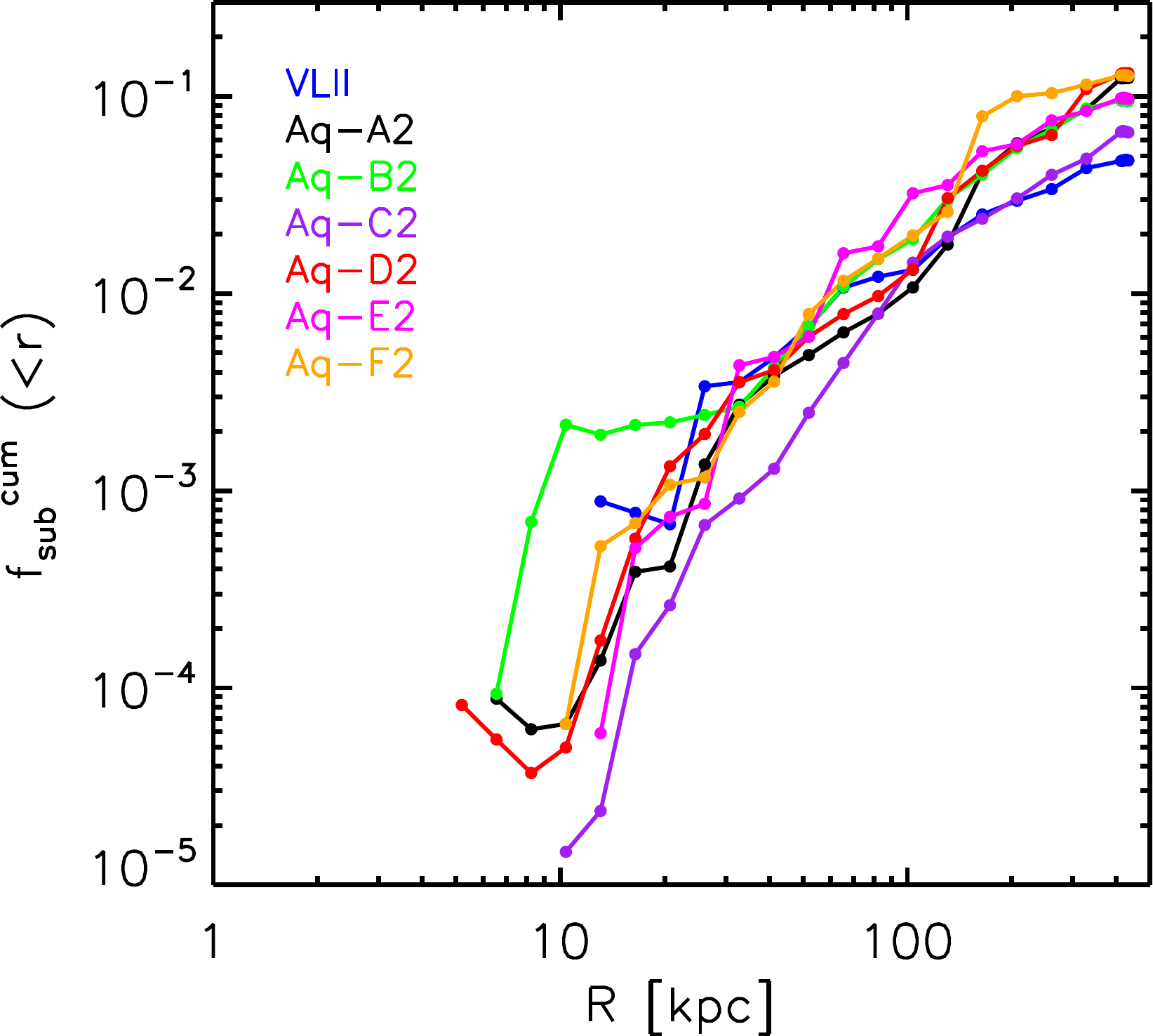}
\caption{Cumulative subhalo mass fraction, substructure mass relative to the halo's enclosed
mass at r, as function of distance from the host halo centre.}
\label{fig:clump}
\end{figure}

\begin{figure}
\includegraphics[width=\linewidth]{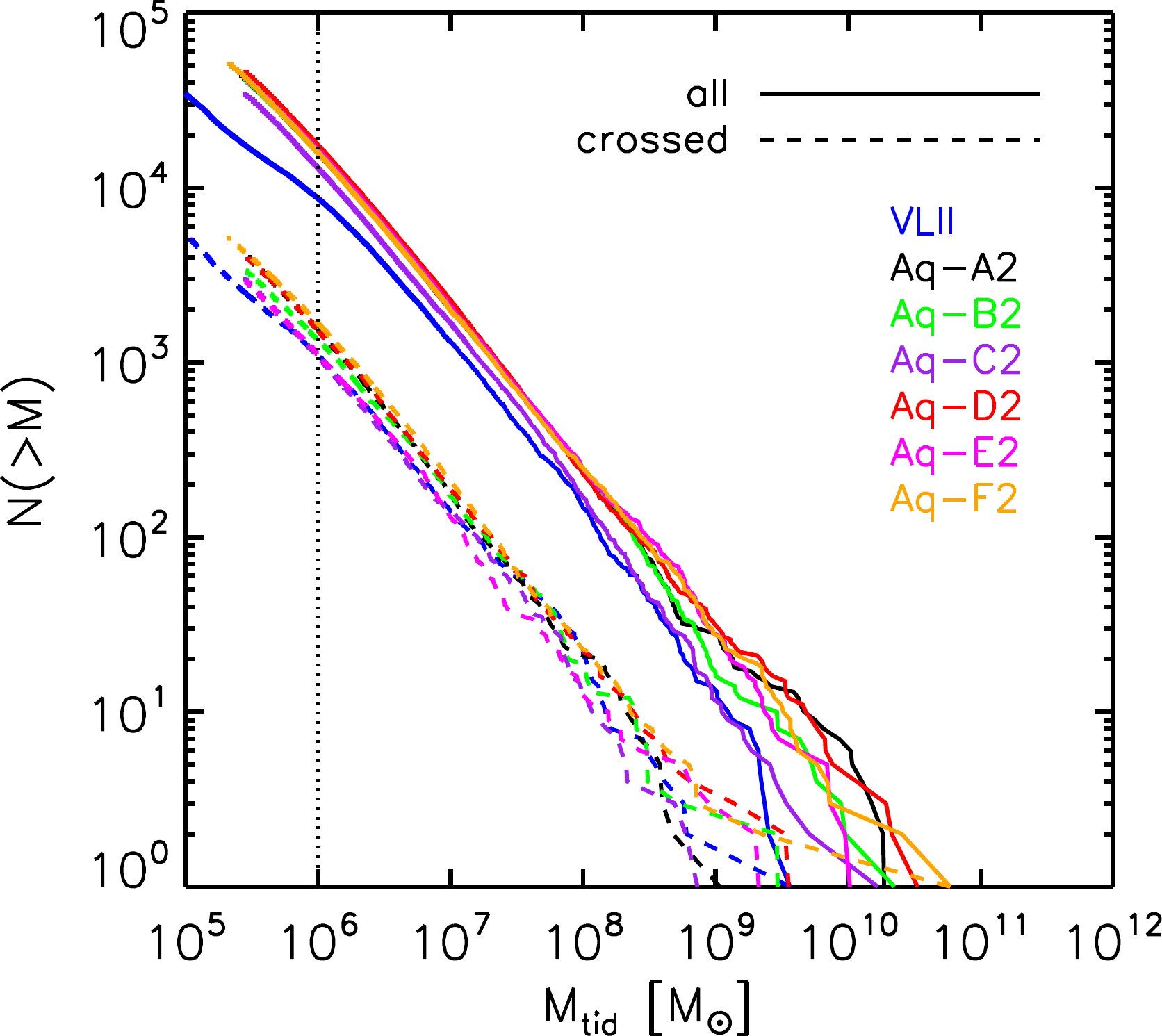}
\caption{Inverse cumulative histogram for the number of subhaloes as function of subhalo mass within $r_{50}$ of the host haloes. The crossed population corresponds to the subhaloes which come closer than 25\,kpc to the centre of the host halo within 2 Gyr while all represents the total sample. The vertical line marks the mass cut at 10$^{6}$ $M_{\odot}$.}
\label{fig:N}
\end{figure}

Fig.\,\ref{fig:clump} illustrates $f_\textrm{sub}^{cum}(r)$, which is the substructure mass relative to the halo enclosed mass at $r$, as function of distance from the parent halo's centre for all seven simulations. Within the inner 20\,kpc Aq-B2 has the highest subhalo mass fraction, while Aq-C2 possesses the lowest value across nearly all radii. Fig.\,\ref{fig:N} illustrates an inverse cumulative histogram for the number of subhaloes as function of their normalised $M_\textrm{tid}$. All the simulations have a similar slope while the scatter in the high-mass regime is the result of low number statistics. For VL\rom{2} the slope decreases at a subhalo mass $\lesssim$ 2 $\times$ 10$^{6}$ $M_{\odot}$; below this mass-range the difference is sensitive to the subhalo finding algorithm and also the mass resolution. Accordingly, the substructure distribution is influenced by lower mass subhaloes that could dominate over higher mass ones by few orders of magnitude in abundance.   
\begin{table*}
\caption{Normalised main halo properties; m$_\textrm{p}$ is the original particle mass and \textit{g} is the scaling factor. $M_\textrm{200}$ is the normalised mass enclosed within a sphere of radius $r_\textrm{200}$, where the mean density of the halo is 200 times the critical density of the universe. c is the concentration while V$_\textrm{max}$ is the maximum circular velocity of the halo. $r_\textrm{50}$ is the radius of the main halo at which the halo encloses the mass with mean density 50 times the \textit{background} density and V$_\textrm{50}$ is the circular velocity at this radii. $f_\textrm{sub}$ represents the cumulative mass fraction which resides in substructures relative to the total enclosed mass within $r_\textrm{50}$.} 
\label{tab:2}
\begin{tabularx}{\textwidth}{@{}lXXXXXXXXr}
 \hline
Simulation & m$_\textrm{p,orig}$ & \textit{g} & $M_\textrm{200}$  & $r_\textrm{200}$ & c & V$_\textrm{max}$ & $r_\textrm{50}$ & V$_\textrm{50}$ & $f_\textrm{sub}$ \\
 & [$M_{\odot}$] &  & [10$^{12}$ $M_{\odot}$] & [kpc] & & [km\,s$^{-1}$] & [kpc] & [km\,s$^{-1}$] & [\%] \\ \hline
VL\rom{2} & 4.10 $\times$ 10$^{3}$ & 1.188 & 1.774 & 242.8 & 13.29 & 212.9 & 425.7  & 151.8 & 4.77 \\
Aq-A2 &  1.37 $\times$ 10$^{4}$ & 0.963 & 1.774 & 242.8 & 16.19 & 205.9 & 428.2  & 156.3 & 12.46 \\
Aq-B2 &  6.45 $\times$ 10$^{3}$ & 2.166 & 1.774 & 242.8 & 9.72 & 204.0 & 418.0  & 152.6 & 9.61 \\
Aq-C2 & 1.40 $\times$ 10$^{4}$ & 1.000 & 1.774 & 242.8 & 15.21 & 222.4 & 417.1  & 152.3 & 6.67 \\
Aq-D2 & 1.40 $\times$ 10$^{3}$ & - & 1.774 & 242.8 & 9.37 & 203.2 & 425.7  & 158.1 & 13.14 \\
Aq-E2 & 9.59 $\times$ 10$^{3}$ & 1.497 & 1.774 & 242.8 & 8.26 & 204.8 & 421.3  & 153.8 & 9.85 \\ 
Aq-F2 & 6.78 $\times$ 10$^{3}$ & 1.564 & 1.774 & 242.8 & 9.79 & 196.3 & 424.7  & 155.0 & 12.80 \\ \hline
\end{tabularx}
\end{table*}
\subsection{Identifying the most effective subhalo candidates}
\label{sec:identify}
As mentioned above, we are interested in subhaloes which are likely to interact with the galactic disc; hence we need to identify such subhaloes. We apply a simple selection criterion by estimating the pericentre distances of all subhaloes. We select these by determining the orbit of subhaloes in the presence of the background potential of the host DM halo only. Using the initial 6-dimensional phase-space information of the subhaloes we integrate their orbits using the standard Runge-Kutta 4th order orbit integrator, which determines the position and velocity of subhaloes at each integration step\,\citep{press}, for 2 Gyr. Any subhalo that comes closer than 25\,kpc to the parent halo's centre at any point during its orbit is marked as a \textit{crossed} candidate. This calculation does not take into account dynamical friction and the gravitational force of the disc and bulge.

\begin{figure}
\includegraphics[width=\linewidth]{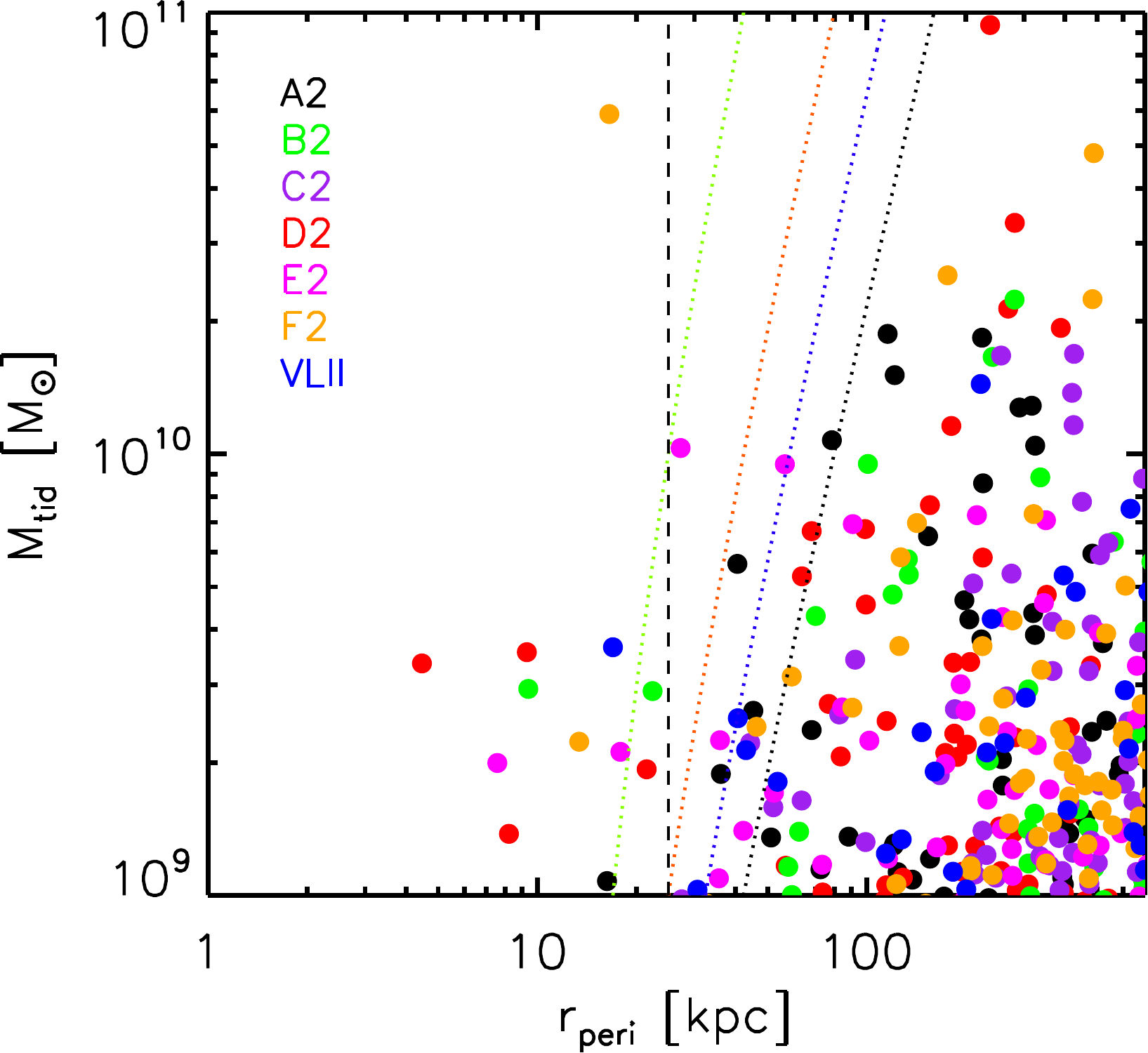}
\caption{$M_\textrm{tid}$ against pericentre distance $r_\textrm{peri}$ for all 7 simulations (colour coded). The vertical dashed line marks the 25 kpc crossing criteria radii; all the subhaloes to the left of this line are regarded as crossed candidates. The 4 dotted lines correspond to lines of same tidal force at 10 kpc from the galactic centre as objects with $M_\textrm{tid}$ of 1, 3, 10 and 100 $\times$ 10$^{8} M_{\odot}$ (from right to left) at $r_\textrm{peri}$=25\,kpc. All subhaloes to the left of these lines have higher tidal forces.}
\label{fig:mtid_peri}
\end{figure}

Fig.\,\ref{fig:mtid_peri} shows $M_\textrm{tid}$ vs. $r_\textrm{peri}$ for subhaloes at the high mass end, with $M_\textrm{tid} > 10^{9} M_{\odot}$, of all simulations (colour coded). The vertical dashed line marks the 25 kpc radius and all the objects to the left of this line are our crossed objects. Instead of using a simple distance cut (and a mass cut at  $M_\textrm{tid} > 10^{8} M_{\odot}$), the strength of the tidal forces on the disc and resonances of the orbital motion and the disc could have been used. The four dotted lines correspond to lines of constant tidal force at 10 kpc from the galactic centre. These correspond to objects with $M_\textrm{tid}$ of 1, 3, 10 and 100 $\times$ 10$^{8} M_{\odot}$ (from right to left) at $r_\textrm{peri}$=25\,kpc. All subhaloes to the left of these lines have higher tidal forces. There is one object in Aq-E2 with a mass of $\sim$ 10$^{10} M_{\odot}$ and a pericentre at 28 kpc, to the left of the red line (10$^{9} M_{\odot}$) which has a tidal force greater than a subhalo with $M_\textrm{tid}$=10$^{9} M_{\odot}$ at $r_\textrm{peri}$=25\,kpc and should have been included; however there are 2 subhaloes in Aq-E2 with larger tidal force (left of green line). It turned out, that only satellites left of the green line contribute to the disc heating on a measurable level. We conclude that we did not miss a significant fraction of interactions, which would change the derived dynamical heating of the disc.

Previous analytical studies proposed a linear scaling between the heating of the galactic disc due to satellites and the subhalo mass $M_\textrm{sat}$. However\,\citet{hopkins} investigated the importance of such dependence using "live" components for disc and subhaloes together with realistic non-circular orbits of satellites. They concluded that the heating rate per Gyr due to continuous infall, which forms the basis of our study, scales non-linearly as $\propto$ $M_\textrm{sat}^{2}$ taking into account dynamical friction. This result contradicts  earlier work which had only focused on the single interaction of subhaloes with discs and scaled linearly as $\propto$ $M_\textrm{sat}$ per merging event. In other words, most of the impact is expected from the most massive subhaloes found in the scatter in the high-mass end of the subhalo mass spectrum (Fig.\,\ref{fig:N}). 

In this study we focus on the interaction of subhaloes with $M_\textrm{tid}$ $\geqslant$ 10$^{8}$ $M_{\odot}$ corresponding to $M_\textrm{tid}$/$M_\textrm{disc}$ $\gtrsim$ 0.003. Table\,\ref{tab:3} includes information regarding the subhalo distribution in all seven simulations together with statistics of the ``crossed" sub-sample. N$_\textrm{sub}$ is the total number of subhaloes in each simulation suite with $M_\textrm{tid}$ $\geq$ 10$^{6}$ $M_{\odot}$, while $p_\textrm{sub}$ shows the fraction of subhaloes within $r_\textrm{50}$. The percentage fraction which cross the introduced 25\,kpc sphere is represented by $p_\textrm{cross}$. VL\rom{2} and Aq-F2 have the largest $p_\textrm{sub}$ values together with the highest crossed fractions. However, Aq-E2 possesses the least number of subhaloes that cross the disc, although it has the fourth highest substructure fraction inside $r_\textrm{50}$. The last four columns of table\,\ref{tab:3} show the number of crossed subhaloes with masses $\geqslant$ 1, 3, 5 and 10 $\times$ 10$^{8}$ $M_{\odot}$. We found a range of subhaloes from 12 to 24 with $M_\textrm{tid}$ $\geqslant$ 1 $\times$ 10$^{8}$ $M_{\odot}$ which provides a good sample for analysing the impact of both the number and mass range of satellites. Aq-D2 has the largest number of subhaloes with mass $\geqslant$ 1 $\times$ 10$^{9}$ $M_{\odot}$ and Aq-F2 is the only simulation that has one candidate $\geqslant$ 1 $\times$ 10$^{10}$ $M_{\odot}$.        
\begin{table*}
\caption{Number statistics of subhaloes; $p_\textrm{sub}$ is the percentage of the subhaloes originally within $r_\textrm{50}$ while $p_\textrm{cross}$ shows the percentage of subhaloes inside $r_\textrm{50}$ which cross the disc in 2 Gyr.
{\large $\star$:} One of the subhaloes has M $>$ 10$^{10}$ $M_{\odot}$} \label{tab:3}
\begin{tabularx}{\textwidth}{@{}lXXccccc}
 \hline
Simulation & N$_\textrm{sub} $ & $p_\textrm{sub}$ & $p_\textrm{cross}$ &  N$_\textrm{cross}$ & N$_\textrm{cross}$ & N$_\textrm{cross}$ & N$_\textrm{cross}$ \\ 
 & &[\%] & [\%] & ($>$ 10$^{8}$ $M_{\odot}$) &  ($>$ 3 $\times$ 10$^{8}$ $M_{\odot}$) & ($>$ 5 $\times$ 10$^{8}$ $M_{\odot}$) &  ($>$ 10$^{9}$ $M_{\odot}$) \\ \hline
VL\rom{2} & 28,618 & 30.36 & 12.71 & 21 & 5 & 3 & 1\\
Aq-A2 & 108,396 & 14.72 & 9.70 & 20 & 6 & 1 & 1 \\
Aq-B2 & 133,793 & 11.85 & 8.45 & 17 & 6 & 3 & 2 \\
Aq-C2 & 121,334 & 10.68 & 8.39 & 14 & 3 & 2 & 0 \\
Aq-D2 & 72,380 & 24.01 & 8.85 & 23 & 8 & 5 & 4 \\
Aq-E2 & 93,322 & 17.49 & 6.77 & 12 & 5 & 5 & 2 \\ 
Aq-F2 & 51,985 & 30.54 & 10.49 & 24 & 9 & 5 & 2$^{\star}$\\ \hline
\end{tabularx}
\end{table*} 

\begin{figure*}
\includegraphics[width=\linewidth]{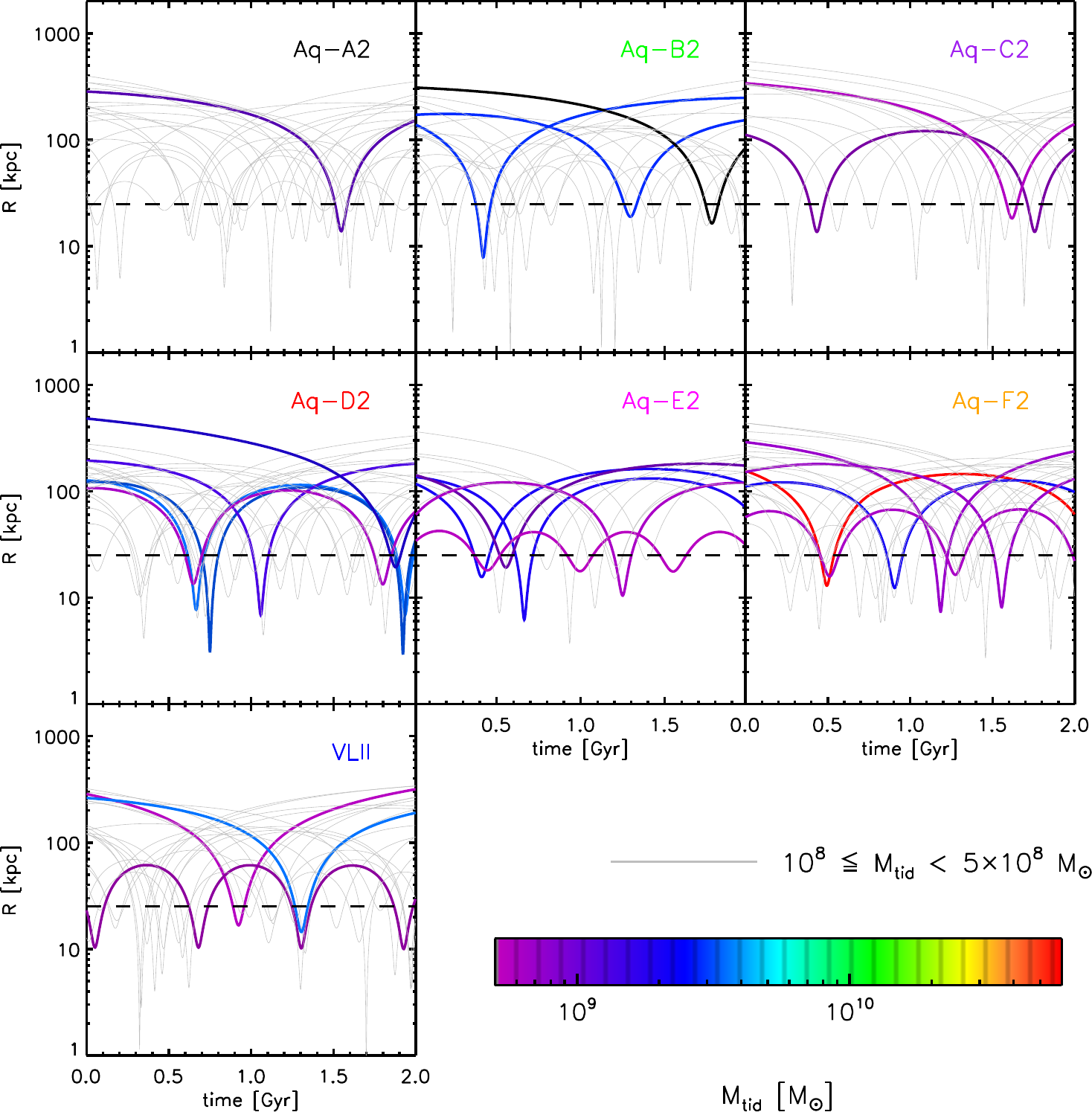}
\caption{The time evolution of the orbits of subhaloes from our N-body simulations in spherical coordinates for all seven host haloes. Subhaloes with 10$^{8}$ $\le$ $M_\textrm{tid}$ $<$ 5 $\times$ 10$^{8} M_{\odot}$ are shown using grey colour while objects with $M_\textrm{tid} \ge 5 \times 10^{8} M_{\odot}$ are represented following a colour map.}
\label{fig:orbit-all}
\end{figure*}

The time evolution of the satellites' orbits from our N-body simulations are shown in Fig.\,\ref{fig:orbit-all} for all 7 simulations. The colours correspond to the mass range of the satellites; 10$^{8}$ $\le$ $M_\textrm{tid}$ $<$ 5 $\times$ 10$^{8}$ (grey) and $M_\textrm{tid} \ge 5 \times 10^{8} M_{\odot}$ following a colour map. The dashed lines represent the 25\,kpc crossing radius. The real pericentric distances from our simulations are very close to the estimated pericentres found in the analysis that used an orbit integrator and the background potential of the host halo (section\,\ref{sec:identify}). The presence of additional mass from bulge+disc slightly reduces $r_\textrm{peri}$, hence the effective limit of selected satellites is $<$ 25 kpc. The crossing of satellites is rather continuous with time.  

\subsection{Subhalo statistics}
\label{sec:stats}
\begin{figure}
\includegraphics[width=\linewidth]{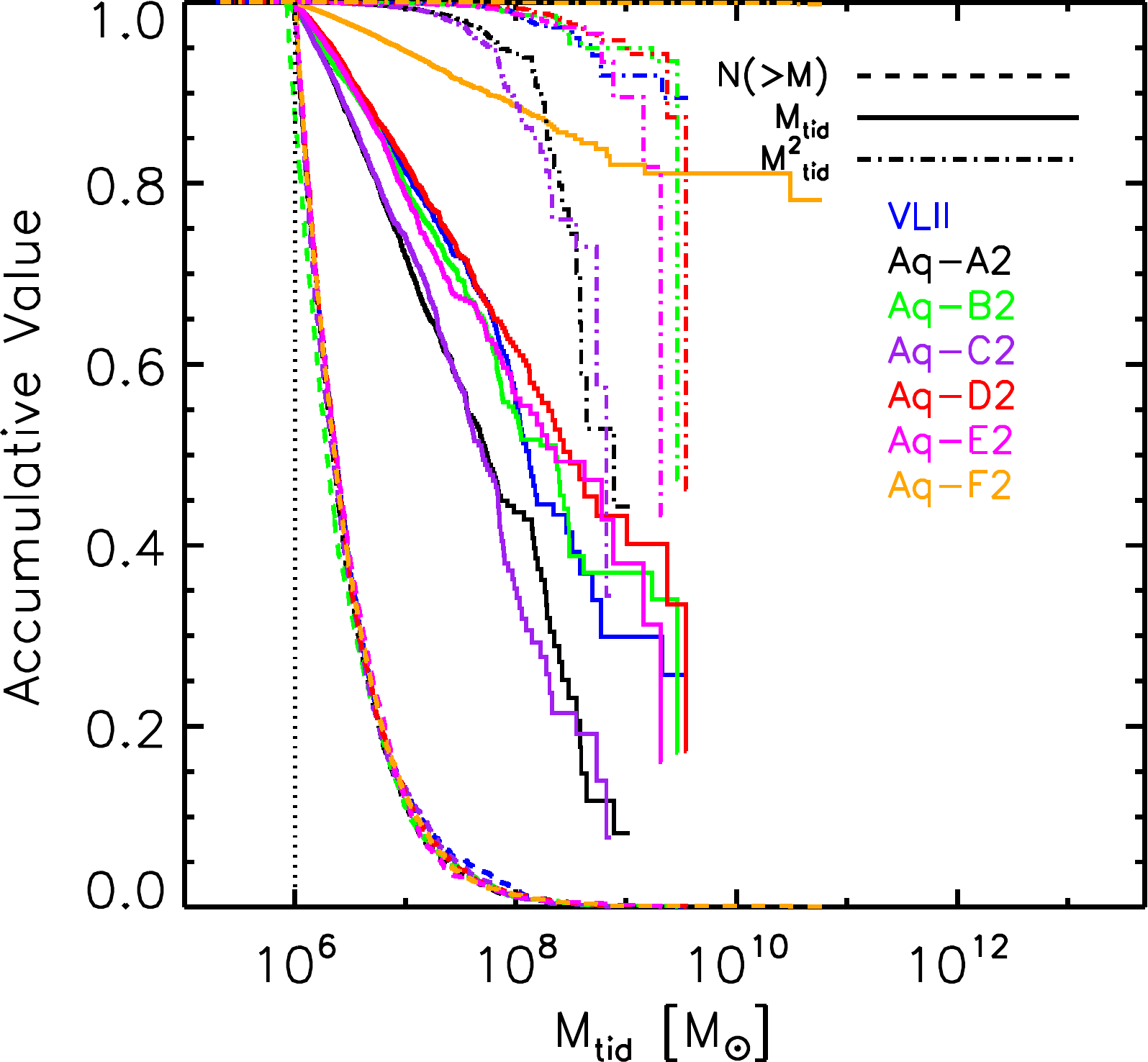}
\caption{Inverse cumulative histogram of number N (dashed), $M_\textrm{tid}$ (solid) and $M_\textrm{tid}^{2}$ (dotted) of crossed subhaloes for all our seven simulations as function of subhalo mass $M_\textrm{tid}$. The vertical line marks the mass cut at 10$^{6}$ $M_{\odot}$.}
\label{fig:normalised}
\end{figure}

Prior to the numerical analysis of the interaction of satellites and the galactic thin disc it is useful to discuss the properties of all subhaloes in our seven cosmological simulation suites. Fig.\,\ref{fig:normalised} represents the inverse cumulative histogram of three quantities: N($>$M) (dashed), $M_\textrm{tid}$ (solid) and $M_\textrm{tid}^{2}$ (dash dotted) for all the \textit{crossed} subhaloes initially within $r_\textrm{50}$. Aquarius simulations have a similar range for N$_\textrm{sub}$. Aq-F2 shows a significant jump at the high mass end of $M_\textrm{tid}$ due to the presence of one subhalo with $M_\textrm{tid}$\,=\,5.9 $\times$ 10$^{10}$ $M_{\odot}$ which actually resides in the minor merger range. For Aq-F2 80\% of the mass resides above 10$^{9}$ $M_{\odot}$, while for the rest of the simulations the same percentage is reached at 10$^{7}$ $M_{\odot}$. As mentioned above the expected impact strength scales with $M_\textrm{tid}^{2}$; more than 95\% of the impact is expected from the really massive subhalo for Aq-F2. Aq-D2 has the second highest expected impact due to massive subhaloes and Aq-C2 has the lowest. In all cases more than 90\% of the impact is expected from subhaloes above 10$^{8} M_{\odot}$.

\begin{figure}
\includegraphics[width=\linewidth]{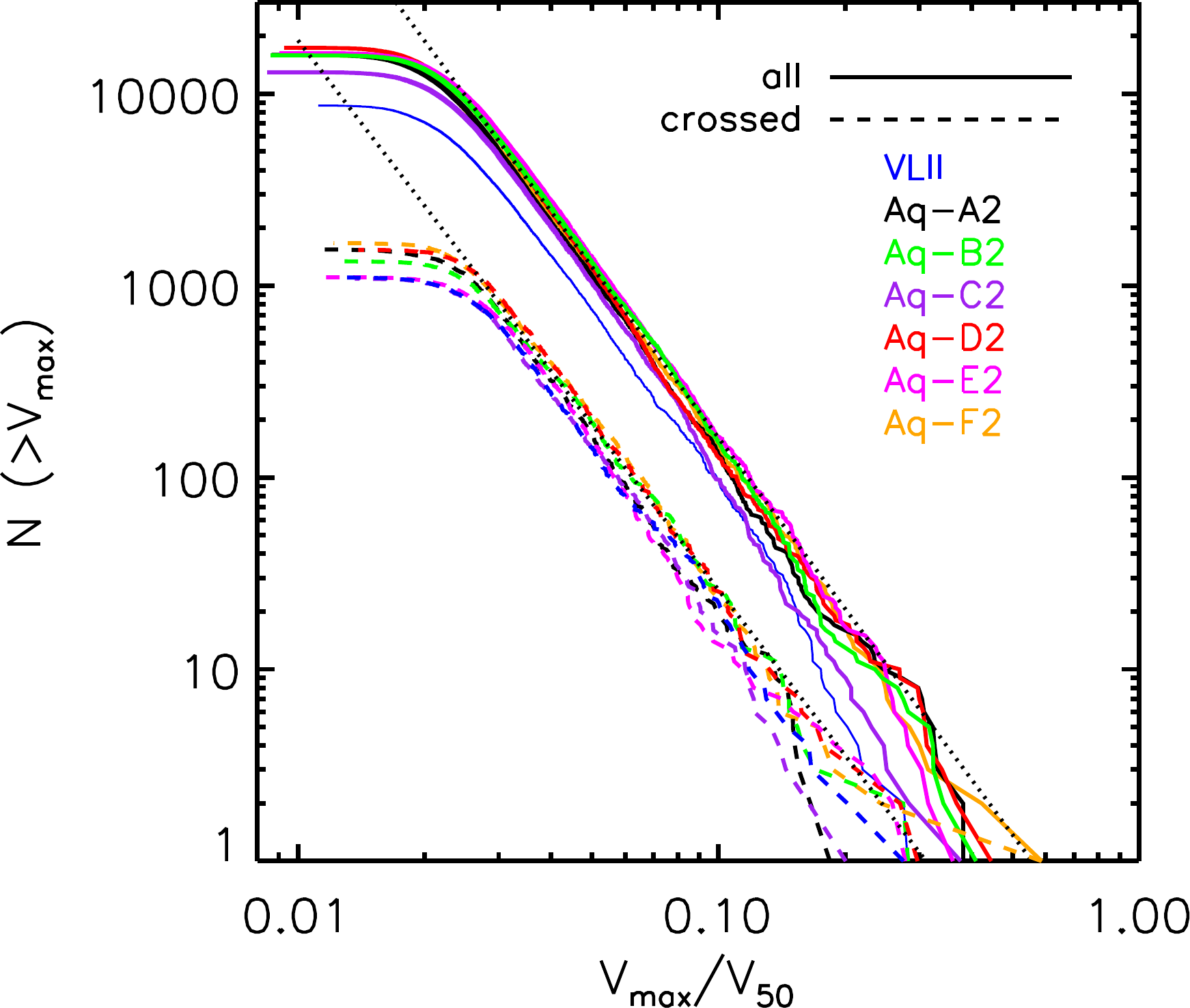}
\caption{Cumulative V$_\textrm{max}$/V$_\textrm{50}$ histogram for all (solid) and crossed (dashed) subhaloes; here V$_\textrm{max}$ is the maximum circular velocity of the subhalo while, V$_{50}$ is the circular velocity of the main halo at $r_{50}$.}
\label{fig:vmax_cum}
\end{figure}

The cumulative histogram of V$_\textrm{max}$ for all (solid) and crossed (dashed) subhaloes is shown in Fig.\,\ref{fig:vmax_cum}. The V$_\textrm{max}$ values are normalised using the circular velocity (V$_\textrm{50}$) of the host halo at $r_\textrm{50}$. About 90\% of the subhaloes have V$_\textrm{max} < 0.05$V$_\textrm{50}$ since the simulations are dominated by lower mass subhaloes. Moving towards subhaloes with V$_\textrm{max} > 0.1$V$_\textrm{50}$, the scatter between the simulations increases due to low number statistics. For all the subhaloes, VL\rom{2} lies on average below Aquarius over the whole spectrum. For the crossed subhaloes it appears the deviation between VL\rom{2} and Aquarius becomes less significant. The two dashed lines correspond to the average of the best-fitting power laws over all the seven simulations except in the case of all the subhaloes were we have excluded VL\rom{2}. The fitted power laws have the form
\begin{equation}
N(>\textrm{V}_\textrm{max})=\beta\, (\textrm{V}_\textrm{max}/\textrm{V}_\textrm{50})^{\alpha}
\label{eq:vmax_fit}
\end{equation}
  \begin{eqnarray}
    \beta &=& 0.176 \times 10^{5} \qquad \alpha=-2.96 \quad \text{all} \nonumber \\ 
    \beta &=& 0.035 \times 10^{4} \qquad \alpha=-2.87 \quad \text{crossed} \nonumber .
  \end{eqnarray}
The slope of the power laws are similar, indicating that the crossed subhaloes are representative of the total population. Also the slope agrees well with that mentioned in\,\citet{aquarius}.
\begin{figure}
\includegraphics[width=\linewidth]{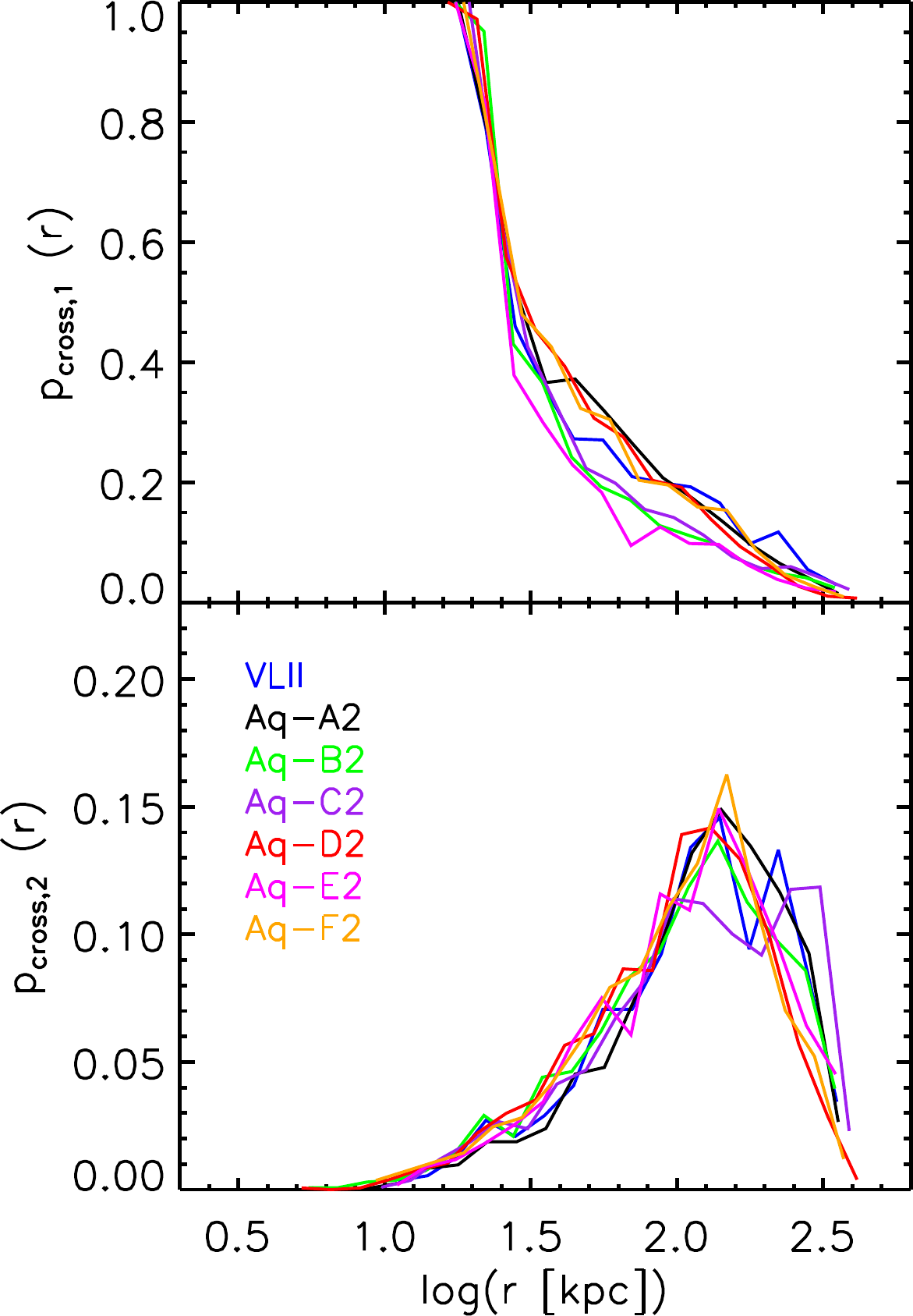}
\caption{\textit{Top}: Fraction of crossed to all subhaloes $p_\textrm{cross,1}$ for spherical shells from centre of their host halo, defined as $\textrm{N}_\textrm{cross}(r)/\textrm{N}_\textrm{all}(r)$. \textit{Bottom}: Normalised radial distribution of all crossed subhaloes, $p_\textrm{cross,2}$, representing $\textrm{N}_\textrm{cross}(r)/\textrm{N}_\textrm{cross}$.}
\label{fig:f_cross}
\end{figure}

Fig.\,\ref{fig:f_cross} presents two quantities $p_\textrm{cross,1}$ (top) and $p_\textrm{cross,2}$ (bottom) which are defined as $\textrm{N}_\textrm{cross}(r)/\textrm{N}_\textrm{all}(r)$ and $\textrm{N}_\textrm{cross}(r)/\textrm{N}_\textrm{cross}$, respectively. The crossed subhaloes located at $r$ $>$ 100\,kpc contribute to less than 20\% of total population at such distance. However these subhaloes make up $\sim$ 50\% of the total crossed fraction $p_\textrm{cross,2}$.    

\begin{figure}
\includegraphics[width=\linewidth]{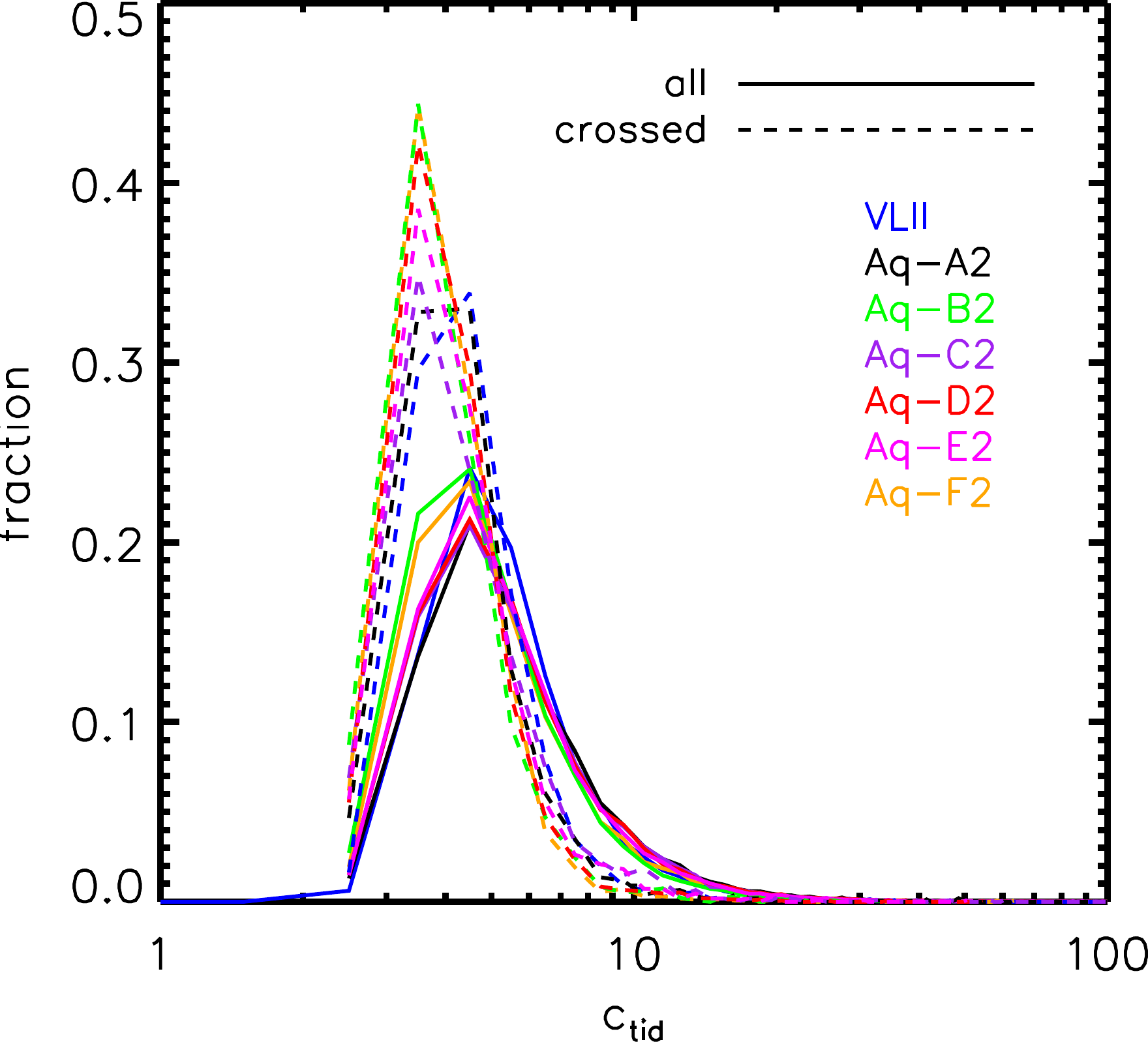}
\caption{Normalised distribution of c$_\textrm{tid}$ for all (solid) and crossed (dashed) subhaloes.}
\label{fig:ctid_hist_norm}
\end{figure}

For the purpose of numerical analysis each crossed subhalo candidate will be generated as a distribution of particles following a NFW profile. To reproduce these subhaloes with a corresponding NFW density profile we determine the radius, $r_\textrm{tid}$, enclosing the provided mass $M_\textrm{tid}$, using the scale radius $r_\textrm{s}$ and V$_\textrm{max}$ of subhaloes. Using 
\begin{equation}
r_\textrm{vmax}=2.163\,\,r_\textrm{s},
\label{eq:rs}
\end{equation}
\begin{equation}	
\left(\frac{\textrm{V}_\textrm{max}}{r_\textrm{vmax}}\right)^{2}=4\,\pi\,\textrm{G}\,\rho_\textrm{sat}(r_\textrm{vmax})
\label{eq:deltav}
\end{equation}
and
\begin{equation}
M_\textrm{tid}=4\,\pi\,\rho_\textrm{s}\,r_\textrm{s}^{3}\,\left[ln(1+\textrm{c}_\textrm{tid}) - \frac{\textrm{c}_\textrm{tid}}{1+\textrm{c}_\textrm{tid}}\right] 
\label{eq:nfw2}
\end{equation}
%
%
\begin{figure}
\includegraphics[width=\linewidth]{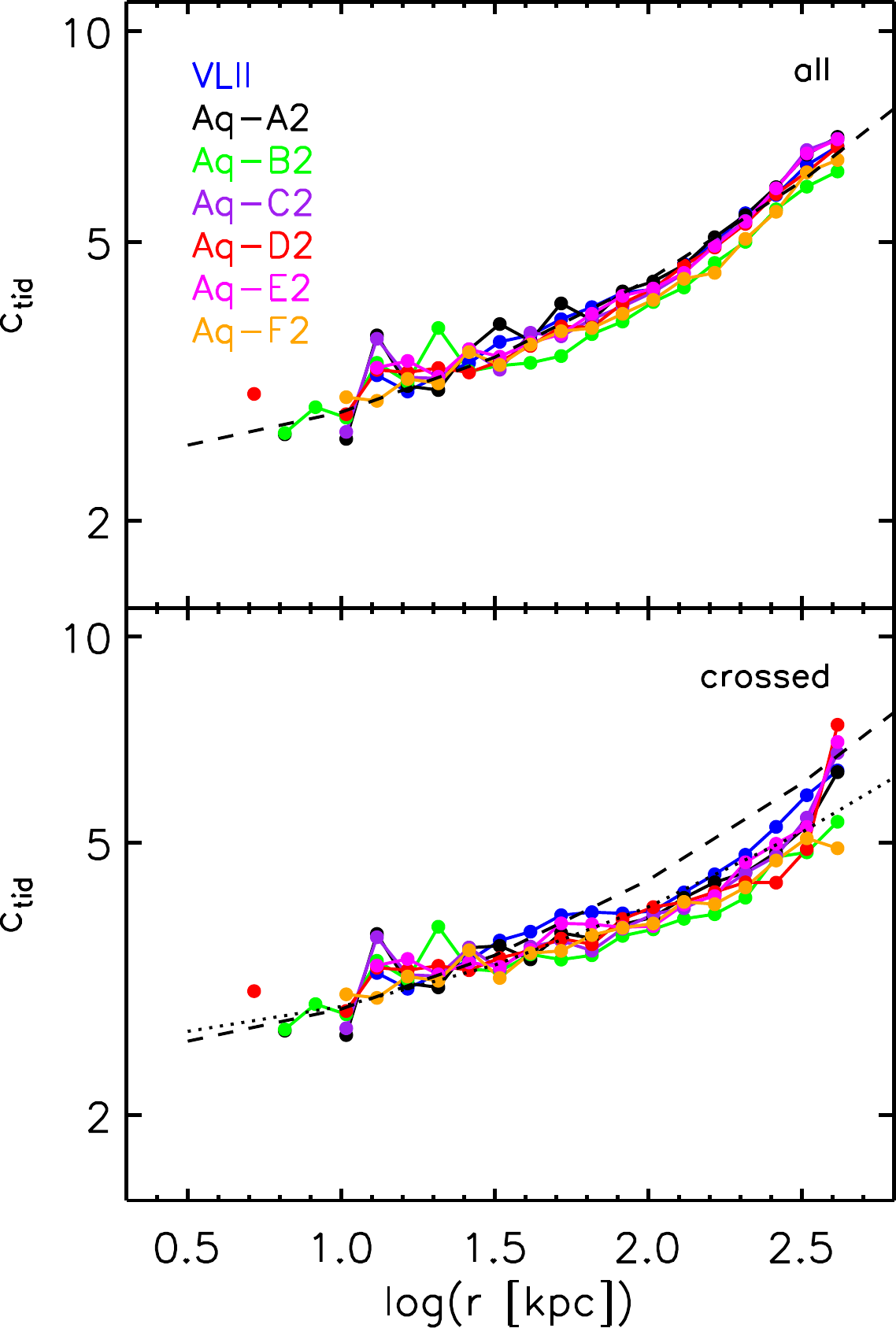}
\caption{Mean concentration c$_\textrm{tid}$ against the distance from centre of the host halo, $r$, for all (top) and crossed (bottom) subhaloes. The dashed and dotted lines represent the average best-fitting power laws for the total and crossed candidates, respectively.}
\label{fig:ctid_r}
\end{figure}
together with equation (\ref{eq:nfw}) outline the most important steps in determining the tidal radius $r_\textrm{tid}$ of subhaloes via calculating c$_\textrm{tid}$ which is the ratio of $r_\textrm{tid}$/$r_\textrm{s}$. The median of the concentration c$_\textrm{tid}$ as a function of distance from the centre of the host halo is shown in Fig.\,\ref{fig:ctid_r} for all (top) and crossed (bottom) subhaloes. The outer most bin at $r \sim r_{50}$ extends to infinity. For guiding the eye, best power law fits for all subhaloes (dashed) and for the crossed ones (dotted) are shown. In both populations we observe higher  c$_\textrm{tid}$ with increasing distance from the host. Hence the subhaloes located further away from the centre tend towards higher concentrations. This trend is due to the tidal mass loss of the subhaloes and is well known\,\citep{nagai}. The scatter between different simulations is insignificant. It is notable that the concentration of  subhaloes located at $r$ $>$ 30\,kpc are, on average, lower for the crossed sample as seen by the deviation of the slope between the fitted power laws. At $r$=150\,kpc where the maximum $p_\textrm{cross,2}$ is observed (Fig.\,\ref{fig:f_cross}), the total population has higher c$_\textrm{tid}$ by a factor of 1.12. For $r$ $<$ 30\,kpc the two populations show very comparable concentrations. This suggests that subhaloes on radial orbits loose more mass, when entering the main halo at $r_{50}$ already. Fig.\,\ref{fig:ctid_hist_norm} is the normalised histogram of c$_\textrm{tid}$ for all (solid) and crossed (dashed) subhaloes. The peak of the distribution for the total population is around c$_\textrm{tid}$ $\sim$ 4.5 while for the crossed subhaloes we observe a shift of the distribution to lower concentrations with the peak around 3.6. This indicates that lower concentration is favoured by crossed satellites. The reason for this behaviour consists of two points. The mean concentration of crossed subhaloes is smaller for $r$ $>$ 100\,kpc  (Fig.\,\ref{fig:ctid_r}) and the fraction of crossed subhaloes, $p_\textrm{cross,1}$, is decreasing with increasing $r$ leading to a biased selection towards lower c$_\textrm{tid}$. 

\begin{figure}
\includegraphics[width=\linewidth]{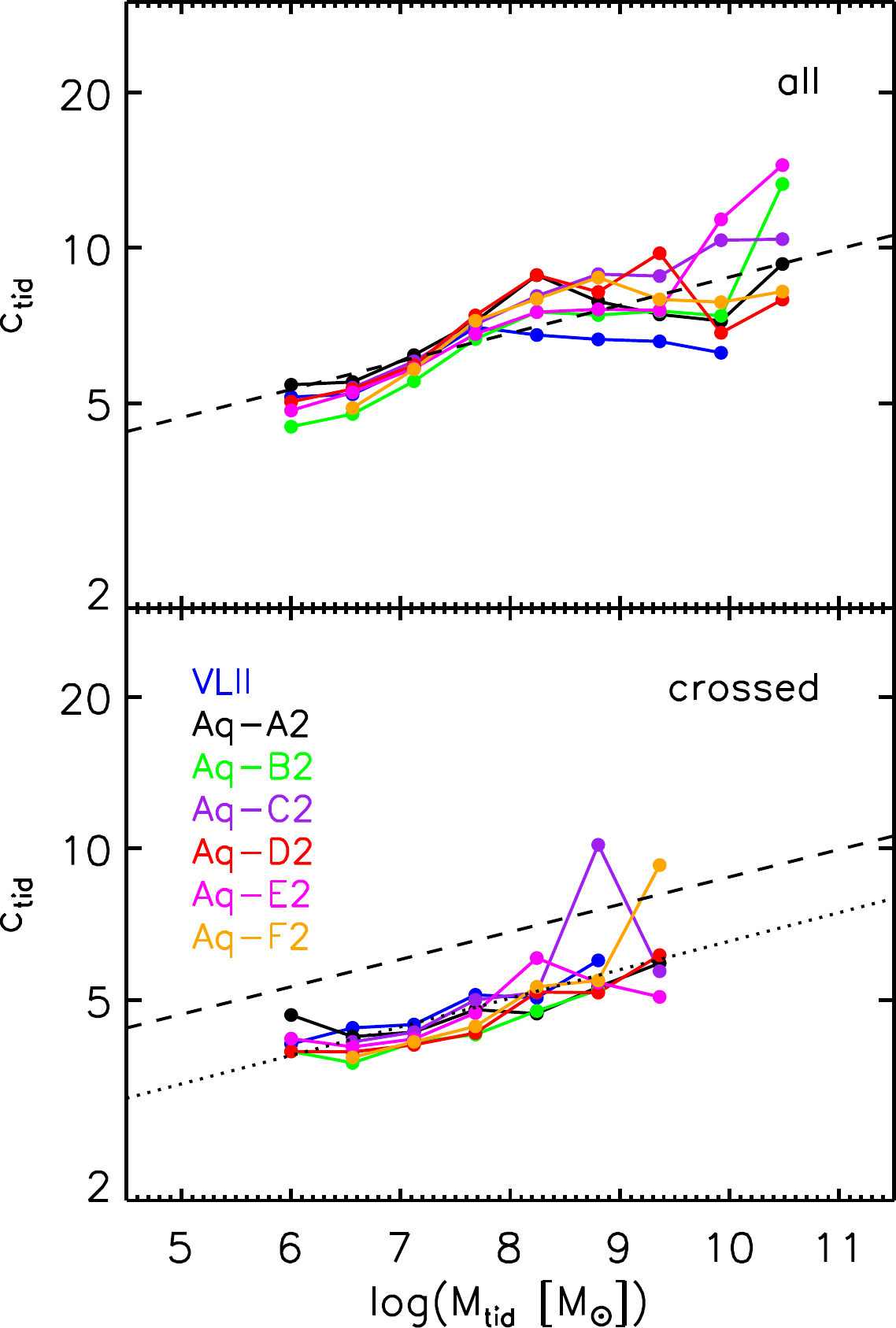}
\caption{Concentration c$_\textrm{tid}$ against $M_\textrm{tid}$ for all (top) and crossed (bottom) subhaloes. The dashed and dotted lines represents the average best-fitting power laws for the total and crossed candidates.}
\label{fig:ctid_m}
\end{figure}

\begin{figure}
\includegraphics[width=\linewidth]{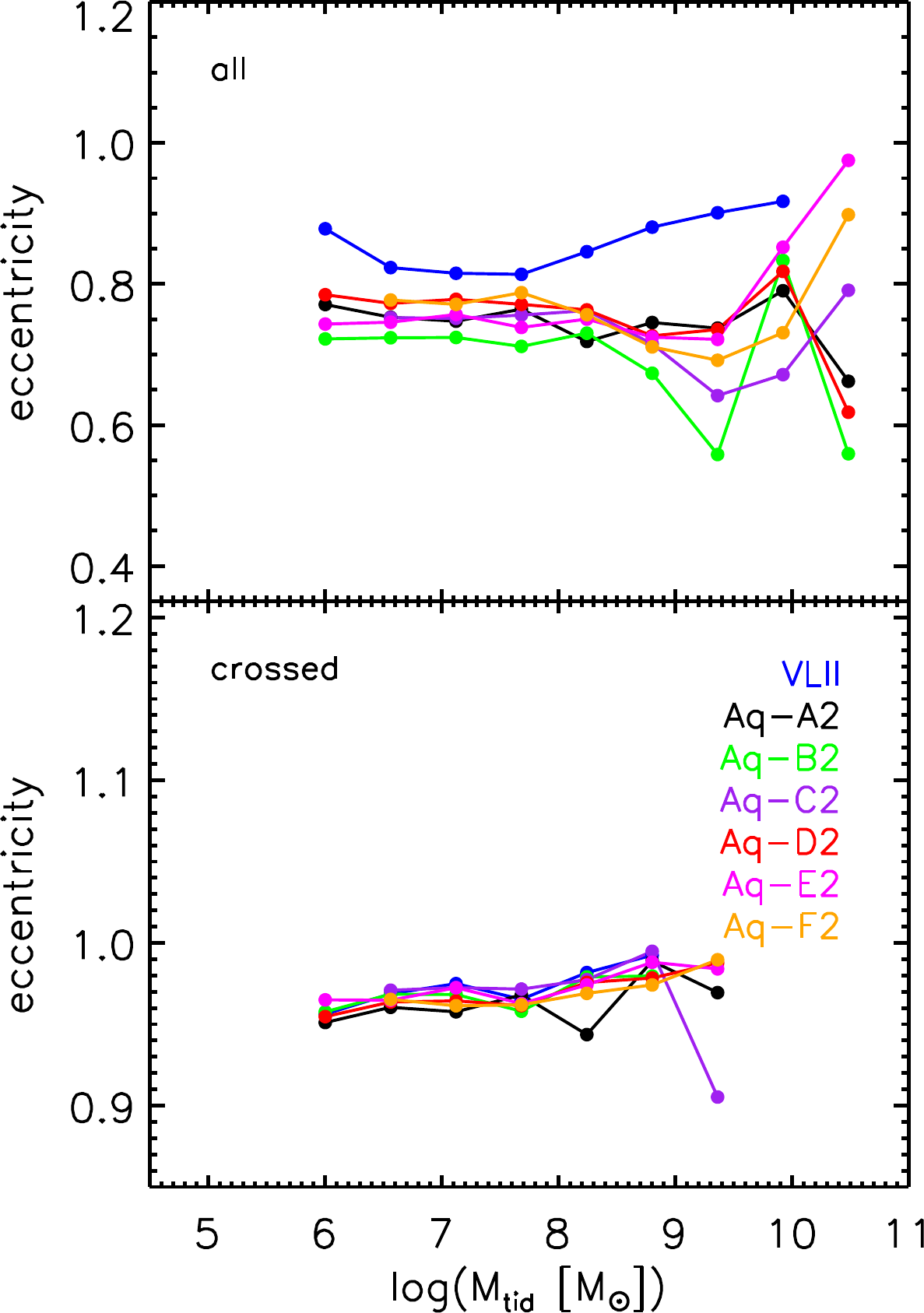}
\caption{Eccentricity \textit{e} against $M_\textrm{tid}$ for all (top) and crossed (bottom) subhaloes.}
\label{fig:ecc}
\end{figure}

The median of the concentration as function of subhalo mass for all (top) and crossed (bottom) subhaloes is plotted in Fig.\,\ref{fig:ctid_m}. Here and also for Fig.\,\ref{fig:ecc} the final bin includes all the subhaloes with larger mass. The scatter for the crossed subhaloes is smaller than for the total population. The dashed and dotted lines represent best fits for all and crossed samples following a power law
\begin{equation}
\textrm{c}_\textrm{tid}=\textrm{c}_{8}\, \left(\frac{\textrm{M}_\textrm{tid}}{10^{8}\textrm{M}_{\odot}}\right)^{\eta},
\label{eq:ctid_fit}
\end{equation}
where c$_{8}$ corresponds to the c$_\textrm{tid}$ value at tidal mass of 10$^{8} M_{\odot}$. The slope of the power law is identical in both plots ($\eta$\,=\,0.05). We observe a shift to lower concentrations for crossed subhaloes by a factor of 1.35 which corresponds to the shifted maximum in Fig.\,\ref{fig:ctid_hist_norm}. The agreement between VL\rom{2} and Aquarius is particularly noticeable for the crossed populations.

\begin{figure}
\includegraphics[width=\linewidth]{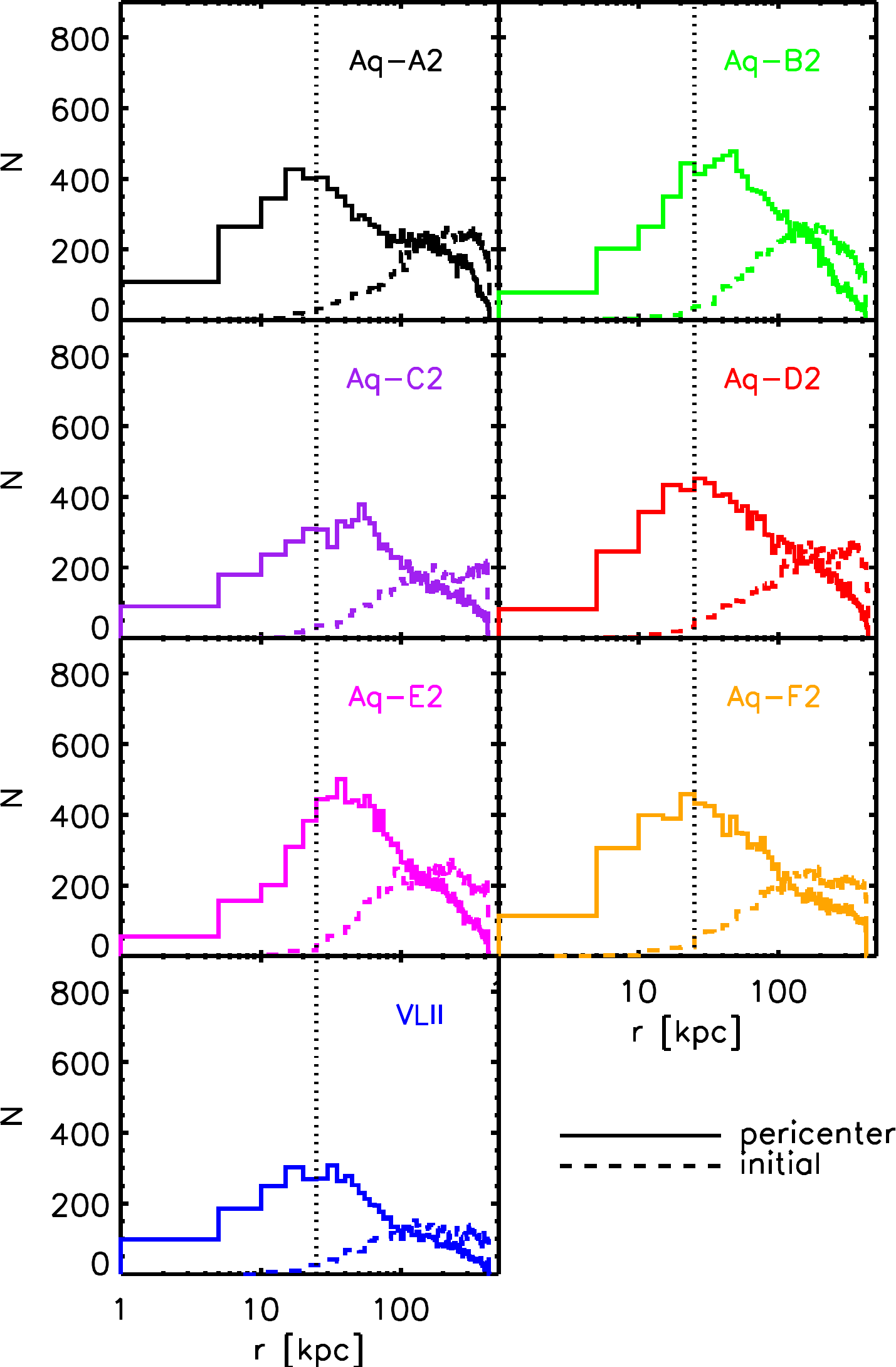}
\caption{Pericentre (solid) and initial distance (dashed) histograms for all subhaloes. The vertical dotted lines mark the 25\,kpc crossing radii.}
\label{fig:peri_initr}
\end{figure}

Fig.\,\ref{fig:peri_initr} represents the distribution of pericentres (solid) and initial distances (dashed) of subhaloes for all seven simulations. The vertical lines mark the position of the 25\,kpc crossing criterion. VL\rom{2} has the lowest number of subhaloes both with pericentres $<$ 25\,kpc and also over the whole distance range. This is partially due to the fact that VL\rom{2} is dominated by low-mass subhaloes, with $M_\textrm{tid}$ $\leq$ 10$^{6}$ $M_{\odot}$, which are ignored in this analysis. There exists some variation in the maximum of the pericentre distribution between the simulations, but most subhalo pericentres lie in the range 10 $<$ $r$ $<$ 100\,kpc.  

\begin{figure}
\includegraphics[width=\linewidth]{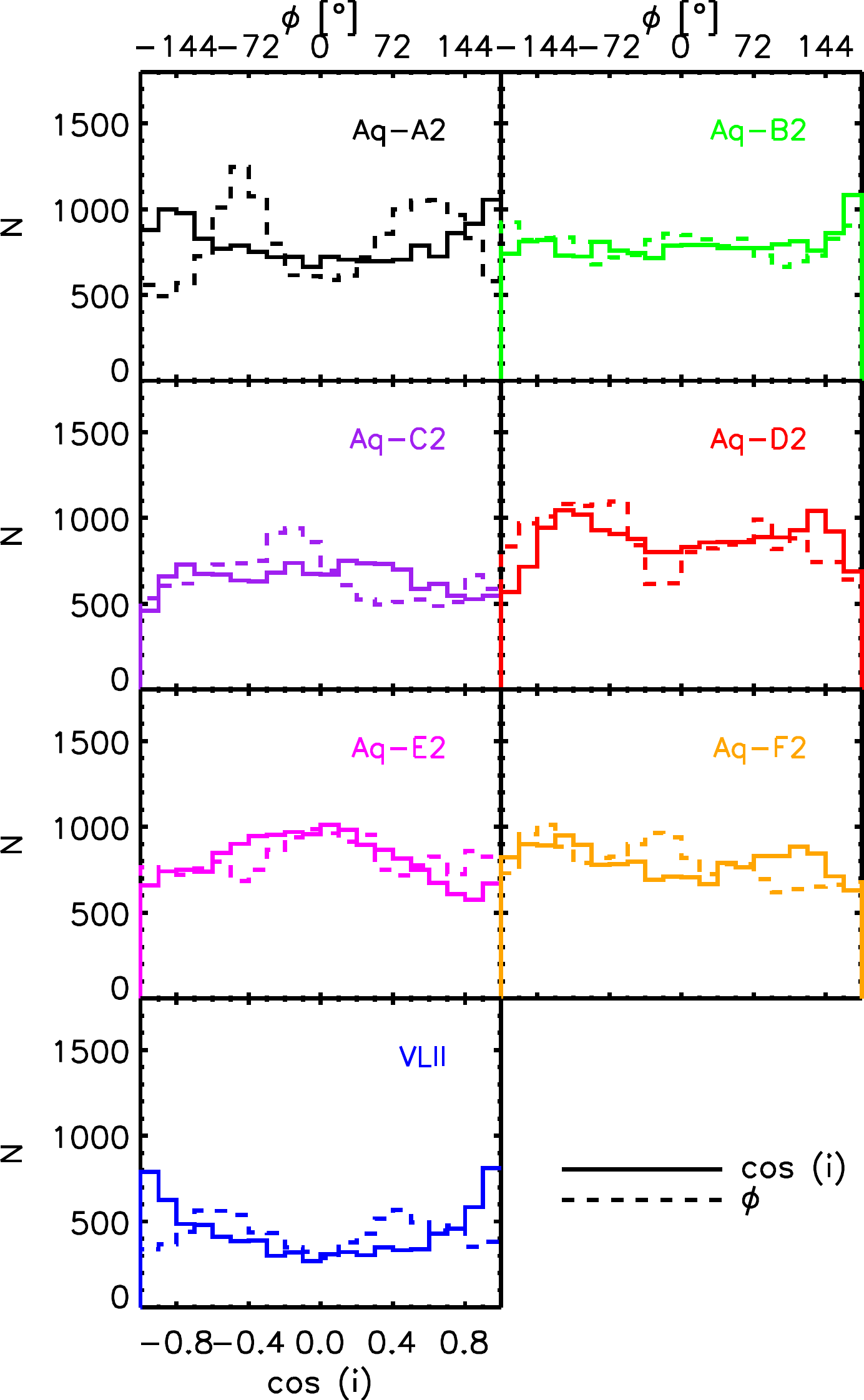}
\caption{Distribution in cos($i$) (solid) and $\phi$ (dashed) for all subhaloes; where $i$ is the subhalo inclination angle with respect to the $z$-component and $\phi$ is the azimuthal angle.}
\label{fig:incl_hist}
\end{figure}

Another interesting aspect is the angular distribution of subhaloes in the simulations. The cosine of the inclination angle \textit{i} (solid) and the azimuthal angle $\phi$ (dashed) of initial positions are represented in Fig.\,\ref{fig:incl_hist} which are measures of \textit{isotropy}. These quantities are calculated by
\begin{equation}
\mathrm{cos}\,(i)=\frac{z}{r} \qquad \textrm{and}
\label{eq:inc}
\end{equation}
\begin{equation}
\phi=\mathrm{tan}^{-1}\left(\frac{y}{x}\right) \qquad -180 < \phi < 180,
\label{eq:phi}
\end{equation}

where $x$,$y$ and $z$ represent the position of the subhaloes in Cartesian coordinates and $r$ is the distance to the host halo's centre in spherical coordinates.

It is important to note in this analysis, the $x$,$y$ and $z$ directions do not necessarily imply any favoured orientation of the system but rather the chosen disc orientation (section\,\ref{sec:model}). In the case of cos($i$), the inclination angle \textit{i} is measured with respect to the $z$-component. Therefore cos($i$) of 1 corresponds to the object which is located on the positive $z$-axis and \textit{vice versa} for cos($i$) of -1. For the Aquarius simulations this quantity has a rather homogeneous distribution with Aq-E2 showing a slight peak around 0, indicating that more subhaloes are concentrated towards the $x$-$y$ plane. However, it is interesting that  in the case of VL\rom{2} we clearly see distinct peaks close to 1 and -1, suggesting that most of the subhaloes are  close to the $z$-axis, i.e. on more polar orbits with respect to the disc. The azimuthal angle $\phi$ tells us how the subhaloes are distributed initially when observed from a top-down view (on the $x$-$y$ plane). We observe the largest deviation from an isotropic distribution for VL\rom{2} and Aq-A2. The bimodal behaviour indicates that more subhaloes are located around $\pm$\,90$^{\circ}$ corresponding to the $y$-axis. Aq-B2 has the most isotropic distribution. We also performed a similar analysis for the crossed subhaloes and concluded that for the Aquarius simulations the distributions for both quantities become more isotropic -- flat in cos$(i)$ and $\phi$ -- while for VL\rom{2} the bimodality is still found. 

One of the important quantities in order to characterize the shape of orbits is the eccentricity \textit{e} which in calculated using
\begin{equation}
e=\sqrt{1-\frac{l^{2}}{\mu a}} \qquad \mathrm{where} \qquad \mu=G(M_\textrm{tid}+M_{h}(r)),
\label{eq:ecc}
\end{equation}
\begin{equation}
l=|\vec{r} \times \vec{v}|,
\label{eq:h}
\end{equation}
\begin{equation}
a=-\frac{\mu}{2\epsilon} \quad \textrm{and} 
\label{eq:semi}
\end{equation}
\begin{equation}
\epsilon=\frac{v^{2}}{2}-\frac{\mu}{r}.
\label{eq:eps}
\end{equation}
In the above equation, $\mu$ corresponds to the sum of the standard gravitational parameters of the subhalo and the host halo with enclosed mass $M_\textrm{h}$ at distance $r$. The specific angular momentum $l$ is calculated using the velocity and position of the subhalo with respect to the host. The semi-major axis \textit{a} and the specific orbital energy $\epsilon$ are related through equation (\ref{eq:eps}). Only in a Kepler potential, the eccentricity and semi-major axis are well-defined and constant along the orbit. A general potential using the enclosed mass, gives an upper limit for the eccentricities and underestimates the pericentre distances.

Fig.\,\ref{fig:ecc} shows the median eccentricity against $M_\textrm{tid}$ for all (top) and crossed (bottom) subhaloes. VL\rom{2} possesses slightly higher \textit{e} values than Aquarius over the whole mass range for the total population. Aquarius simulations have a relatively similar behaviour in the $M_\textrm{tid}$ $<$ 10$^{9}$ $M_{\odot}$ zone, while the scatter in the higher mass regime is expected. For the crossed subhaloes, the difference between the simulations is minimized and there exists a minor increase of \textit{e} with $M_\textrm{tid}$. In other words the orbits of crossed subhaloes become more parabolic, closer to \textit{e}\,=\,1, for higher masses.   

More than 85\% of the subhaloes in all seven simulations have 0.5 $<$ \textit{e} $<$ 1 -- these subhaloes have bound elliptic orbits with respect to the enclosed halo mass at the corresponding radii. Also $\geq$ 10\% have hyperbolic trajectories with \textit{e} $>$ 1 and regarded as candidates which would cross once as they pass by the galactic disc. VL\rom{2} has the highest fraction of subhaloes with unbound orbits.

\begin{figure}
\includegraphics[width=\linewidth]{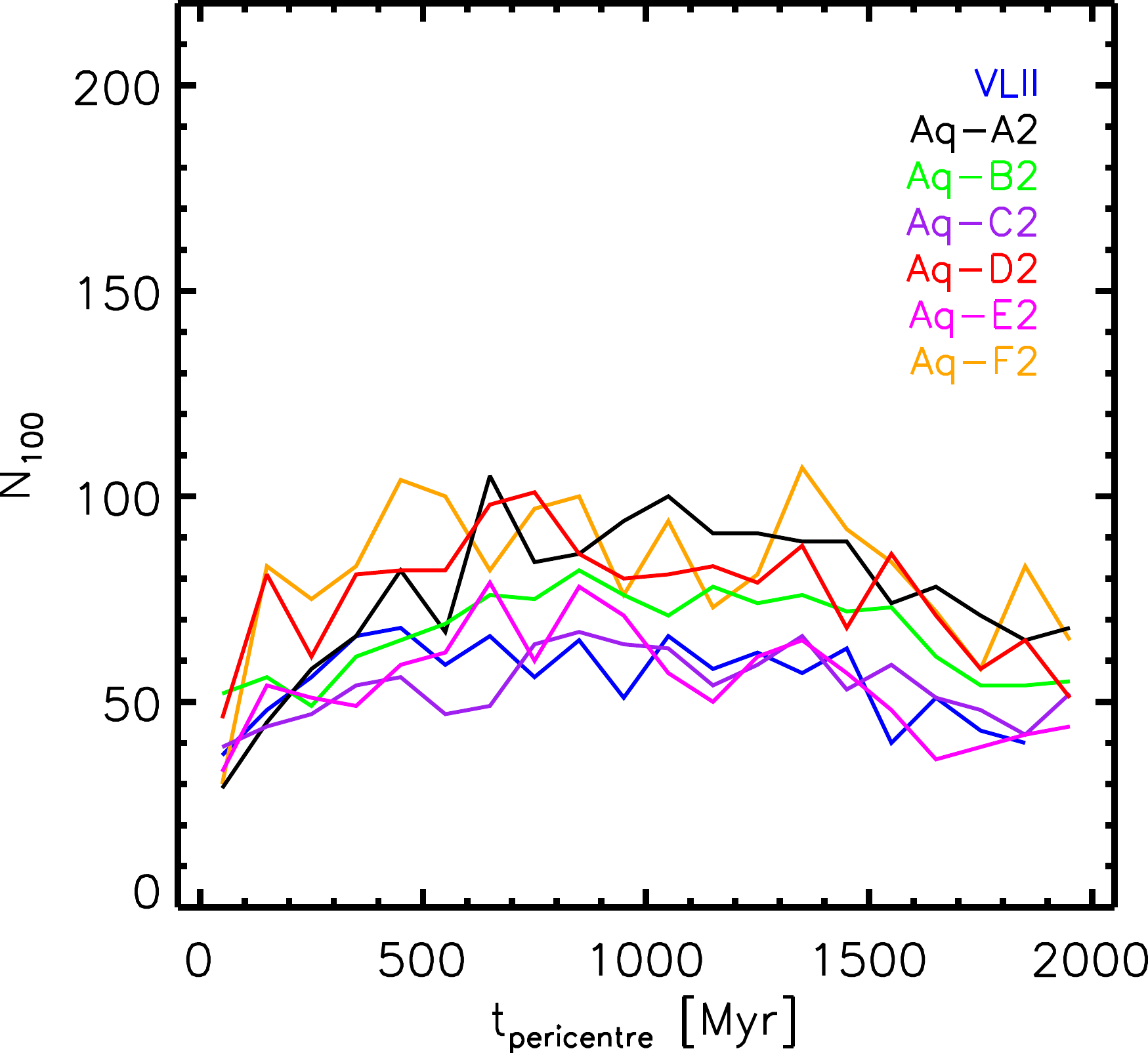}
\caption{Distribution of pericentric time t$_\textrm{pericentre}$, time of reaching the closest pericentre, for crossed subhaloes. Multiple pericentre passages are not included.}
\label{fig:peri-t-hist}
\end{figure}

In order to determine how homogeneous is the infall time distribution of subhaloes, we looked at the pericentric passage time of the subhaloes which correspond to the time of the closest approach to the centre of the host halo. The expected pericentric time of crossed subhaloes for all the seven simulations is shown in Fig\,\ref{fig:peri-t-hist}. The distribution is fairly homogeneous while we observe a shallow peak around 1 Gyr time; hence the impact of the subhaloes is expected to be continuous with time. There is a lack of encounters for the first 0.5 Gyr which might be due to mass loss of subhaloes already located at close distances. The possible impact of this delay on heating of the galactic disc is discussed in section\,\ref{sec:results}. Also the drop seen in all the simulations after 1.5 Gyr could be the result of simulation box's finite volume size. This plot can robustly tell us about the interaction rate per Gyr of subhaloes with the host.

The VL\rom{2} and Aquarius simulations do not show significant differences in their spatial distribution. The higher mean eccentricities of the VL\rom{2} satellites is the main difference compared to the Aquarius simulations. However for the crossed subhaloes such a difference disappears.

\section[]{Galaxy N-body Model}
\label{sec:model}
Prior to the numerical analysis we need to set up our desired galaxy models as a distribution of particles which include disc, bulge, DM halo and also the satellite component. As mentioned earlier, one of the most important aims of this work is to resolve the dynamical properties of the galactic disc in the absence of any bias originating from a system which is not originally in equilibrium.

\subsection{Multi-component Milky Way galaxy}
In the past two decades there has been a number of attempts to generate initial conditions for multi-component spiral galaxies that include a disc, bulge and DM halo component, and producing a distribution of particles which follow particular density profiles. We have tested a few  such attempts, however the true equilibrium and steady state of the disc component is questionable. For this study we take advantage of a recent code (GalIC) which employs a different approach to the made-to-measure method: constructing ICs for multi-component axisymmetric systems via an iterative scheme\,\citep{yurin}.

The stellar disc component has an exponential radial profile while for the vertical direction an isothermal profile is adopted 
\begin{equation}
\rho_\textrm{d}(R,z)=\frac{\textrm{M}_\textrm{d}}{4\pi z_{0}h^{2}}\,\textrm{exp}\left(-\frac{R}{h}\right)\textrm{sech}^{2}\left(\frac{z}{z_{0}}\right)
\label{eq:disc}
\end{equation}
In the above equation \textit{h} and \textit{$z_{0}$} correspond to the radial scale length and the thickness of the disc; the total mass of the disc is represented by $M_\textrm{d}$. For the scaling of the disc surface density, we use the value of the thin disc at the solar neighbourhood $\Sigma_\textrm{sol}$ proposed by\,\citet{just} which is then converted into $M_\textrm{d}$ via
\begin{equation}
\textrm{M}_\textrm{d}=2\pi\Sigma_\textrm{sol}h^{2}\,\textrm{exp}\left(\frac{R_{0}}{h}\right)
\label{eq:surf}
\end{equation}
where R$_{0}$\,=\,8\,kpc is supported by\,\citet{gillessen} and\,\citet{reid}. The value for the radial scale length \textit{h} is chosen to be 2.8\,kpc. Since the disc mass is not sensitive to the scale length over a large range, when fixing the local surface density, a strong influence from smaller scale lengths as discussed in recent literature is not expected. The GalIC code allows us to constrain the kinematical preferences of axisymmetric components. For the disc, we use the ratio of the radial and vertical velocity dispersions $\sigma_{R}/\sigma_{z}$ which converges to a unique solution during the IC construction. The bulge follows a spherically symmetric and isotropic Hernquist profile (equation\,\ref{eq:hern}). The DM halo is set up in two stages. The first step consists of using properties of our DM halo such as \textit{$\lambda$}, c and V$_{200}$ (the spin parameter, concentration and total circular velocity of the system at $r_{200}$) extracted from the seven simulations. These quantities together with the employed cosmology allows us to infer the total mass (disc+bulge+DM halo) of the system, $M_\textrm{tot}$, using
\begin{equation}
\textrm{M}_\textrm{tot}=\frac{\textrm{V}_{200}^{3}}{10\,GH_{0}}.
\label{eq:mtot}
\end{equation}
The spin parameter $\lambda$, which determines the size of the disc, possesses a range of values for different haloes and is dependent on the concentration. Since we would like to have similar disc profiles among all simulations, this quantity is calculated such that the discs have similar scale lengths. These systems lie in the low spin parameter range. The code takes a reverse procedure and using the inputted \textit{$\lambda$} together with the disc mass fraction m$_{d}$, it calculates the scale length of the disc (e.g. \citealt*{mo,springel2}). In the second step, the code uses the technique where a matching Hernquist profile is created for our desired NFW halo. This profile has a similar density profile to NFW in the inner part while falling off steeper in the outer regions according to
\begin{equation} 
  \rho_\textrm{hern}(r)=\frac{M_\textrm{hern}}{2\pi}\frac{a}{r(a+r)^3},
  \label{eq:hern}
\end{equation}
where $M_\textrm{hern}$ is the total mass of the halo and $a$ is the scale radius of the Hernquist profile.\,\citet*{springel2} introduced a method for constructing a Hernquist model with a density profile which matches that of a particular NFW profile. This is achieved by associating a Hernquist DM halo with total mass - $M_\textrm{hern}$ - same as the NFW halo mass within $r_\textrm{200}$ where scale lengths are related via the NFW halo concentration 
\begin{equation}
  a=r_\textrm{s}\sqrt{2[ln(1+c)-c/(1+c)]}.
  \label{eq:match}
\end{equation}
We used this approach to generate the matching Hernquist potential for the parent haloes and employ them in the orbit integration analysis. We also found negligible deviations between the orbits of subhaloes in the NFW and the corresponding Hernquist haloes of the order of $<$ few percent. All the DM haloes in this study are generated with a spherical density profile for simplicity, while the original profile from the cosmological simulations are non-spherical, i.e. prolate/oblate/triaxial,\,\citep{allgood}. 

Some of the quantities used for the set up of the initial conditions are listed in table\,\ref{tab:4}. The $\lambda$ ranges from 0.021--0.028 depending on the simulation and the DM halo which is constructed. Each system contains 10, 0.5 and 4 million particles for the disc, bulge and the halo respectively. According to\,\citet{khoperskov} an insufficient number of disc particles could damp out the modes by introducing noise; hence 10 million disc particles allows for a physically reliable analysis of the disc.
\begin{table}
\caption{The characteristics of the components for the Milky Way. N represents the number of particles for the component while the subscripts d,b and h correspond to the disc, bulge and the halo. The solar neighbourhood surface density $\Sigma_\textrm{sol}$ together with the thin disc scale height and scale radius, \textit{$z_{0}$} and \textit{h} correspond to the model proposed by\,\citet{just}.} \label{tab:4}
\begin{tabularx}{\linewidth}{XX}
 \hline
Quantity & Value \\ \hline
$\Sigma_\textrm{sol}$ & 39.9 $M_{\odot}$pc$^{-2}$\\
$h$ & 2.8 kpc \\
$z_{0}$ & 277 pc \\
$M_{d}$ & 3.42 $\times$ 10$^{10}$ $M_{\odot}$ \\
$\sigma_{R}/\sigma_{z}$ & 1.5 \\
N$_{d}$ & 10 million \\
a$_{b}$ & 0.35 kpc \\
$M_{b}$ & 1.9 $\times$ 10$^{10}$ $M_{\odot}$ \\
N$_{b}$ & 0.5 million \\
$M_{200}$ & 1.77 $\times$ 10$^{12}$ $M_{\odot}$ \\
$\lambda$ & 0.021-0.028 \\
N$_{h}$ & 4 million \\ \hline
\end{tabularx}
\end{table} 
\subsection{Satellite model}
The satellites with $M_\textrm{tid}$ $\geq$ 10$^{8}$ $M_{\odot}$ are generated using their $M_\textrm{tid}$, c$_\textrm{tid}$ and $r_\textrm{s}$ calculated in section\,\ref{sec:stats} together with the distribution function introduced by\,\citet{widrow} and implemented in\,\citet{lora} for a cuspy NFW profile. We need to employ a cut-off radius for the satellites due to the infinite cumulative mass of NFW profile at $r$ $\rightarrow$ $\infty$; hence $r_\textrm{tid}$ was chosen. All the satellites for this study are represented using 50,000 particles. 

\subsection{N-body code}
\label{sec:code}
The numerical simulations in this paper are performed using the efficient parallelised N-body particle-mesh code SUPERBOX-10\,\citep*{bien}. Since the individual gravitational interaction of particles from similar/different components, also known as two-body relaxation, has a time-scale of the order of the Hubble time  we can regard them as collisionless\,\citep{binney}. SUPERBOX-10 has three nested grids at different resolution for \textit{each} galaxy, where the inner most grid with the highest resolution allows the inner dense structure of every galaxy to be resolved and the outer most grid contains the whole simulation and coincides for all galaxies. Each grid is divided into cubic cells and we decided to use $n^3$ cells with $n$=256 instead of 128 in order to reduce the energy conservation error to $<$ 0.4\% after 4 Gyr of evolution. The size of each cell depends on the radius of the grid R$_\textrm{grid,i}$ and calculated as
\begin{equation}
\textrm{d}_\textrm{cell}=\frac{2\,\textrm{R}_\textrm{grid,i}}{n-4}.
\label{eq:cell}
\end{equation}
The outermost grid contains all the particles in the simulation and hence has a size of few hundred kpc. The calculation of the potential is performed via transforming the cell densities into potential using parallelised Fast Fourier Transform (FFT) together with Green's function. The computation time scales as Nlog\,N in the case of FFT while direct calculation scales as N$^2$. In addition, the orbital integration of the particles are done using a leapfrog method and a constant time step\,\citep{superbox}. It is important to choose an appropriate time step as it plays an essential role in the force calculations; it needs to be short enough that a typical particle in the inner most grid moves a maximum of one cell during the step, which was chosen to be 0.1 Myr.

In order to enhance the resolving power of the vertical structure of the galactic disc, SUPERBOX-10 takes advantage of a flattening parameter \textit{q} which flattens the middle grid along the \textit{z}-axis. This technique improves the calculation of the acceleration in the vertical direction. For all our simulations we use similar grid radii for the Milky Way galaxy; R$_\textrm{inn}$\,=\,3.5\,kpc, R$_\textrm{mid}$\,=\,30\,kpc and R$_\textrm{out}$\,=\,500\,kpc.  The cell size of each grid are d$_\textrm{inn}$\,=\,0.028\,kpc, d$_\textrm{mid}$\,=\,0.24\,kpc and d$_\textrm{out}$\,=\,4.0\,kpc. The flattening factor \textit{q} has the value of 0.3, which increases the resolution of the middle  grid in the $z$ direction by a factor of $\sim$ 3.3. According to the work by\,\citet{just11} the effective resolution in the force calculations using the particle-mesh code is d$_\textrm{eff}\,=\,d_\textrm{cell}/2$. Hence we reach a resolution of 40 parsecs in the vertical direction of the entire disc in the middle grid.

Each satellite has its own 3 grids, where the sizes of the inner and middle grids are chosen as 2$r_\mathrm{s}$ and 2$r_\mathrm{tid}$; the outer grid has the same size as the MW system, 500\,kpc.

\subsection{Analysis of the disc}
The disc is subject to being shifted, tilted and distorted with respect to the original coordinate system due to evolution and mostly the bombardment by the satellites. Therefore we defined at each time-step a new coordinate system for the analysis of the disc structure by moving the origin to the density centre of the host galaxy. The velocities of the disc particles are corrected for the density centre velocity, then the coordinate system is rotated to correct for the tilt. It turned out that the angular momentum vector of the disc is the most robust measure of the disc rotation axis. The positive $z$-axis is rotated to the angular momentum vector of the disc, taking into account all disc particles. Since we are interested in the radial range of 4-15\,kpc (for an exponential disc with flat rotation curve 95\% of the disc mass and 87\% of the angular momentum is enclosed within 5\,$h$=14\,kpc). The tilt angle ranges from $<$ 1$^{\circ}$ for most of our simulations to $\sim$ 4$^{\circ}$ in the case of Aq-F2 with satellites.

For the analysis of the disc structure we study the temporal evolution of the solar neighbourhood defined by 7.5$<$R$<$8.5\,kpc, and the radial profile at the end of the simulation. We evaluate in radial rings, $z_\textrm{mean}$ which provides a measure of the mean value of the disc $z$-coordinate, in order to check the disc orientation and identify bending modes. We have checked that the inner disc is well-aligned with the new coordinate system in all cases. The root-mean-square of the disc \textit{z}-coordinate in the solar neighbourhood calculated via    
\begin{equation}
z_\textrm{rms}=\sqrt{<(z-z_\textrm{mean})^{2}>}, \quad |z-z_\textrm{mean}| \leq 2\,\textrm{kpc}
\label{eq:zstd}
\end{equation}
which is a measure of the disc thickening and the square of the vertical velocity dispersion are measures of disc's vertical heating. We neglect any disc particle with $|z-z_\textrm{mean}| \geq$ 2\,kpc for this analysis in order to avoid any outliers.

\subsection{Isolated system}
\begin{figure}
\includegraphics[width=\linewidth]{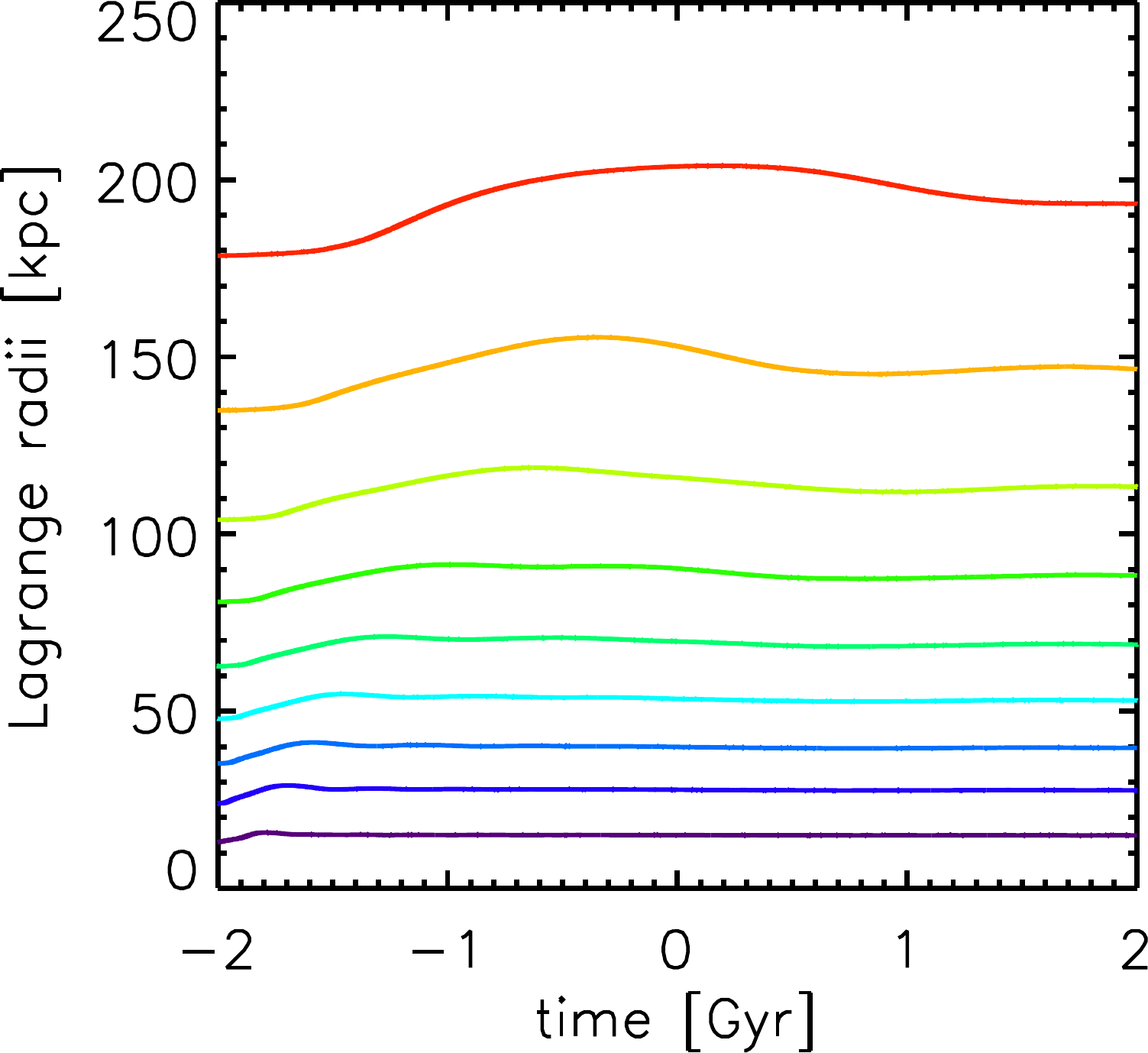}
\caption{10-90\% Lagrange radii of the isolated MW for Aq-B2 simulation. The 10\% radii corresponds to the lowest value (dark purple) while 90\% has the highest value (red) and the remaining colours in between show the 20-80\% Lagrange radii from bottom to top.}
\label{fig:B2-radb}
\end{figure}

The multi-component Milky Way galaxy was evolved in isolation -- in the absence of satellites -- for 2+2 Gyr in the case of all 7 simulations. The initial 2 Gyr is for the purpose of reaching an equilibrium state while the latter 2 Gyr is to have a control group for comparison to the systems with satellites. The isolated evolution allows the components to adapt to their environment and also reach the equilibrium and steady state. Fig.\,\ref{fig:B2-radb} shows the time evolution of 10-90\% Lagrange radii of the Aq-B2 system (disc+bulge+halo) in isolation. This quantity represents the spherical radii at which 10-90\% of the total mass are enclosed. The inner 50\% of the system that extends beyond the disc does not show signs of expansion/contraction during 4 Gyr. The outer regions of the halo have only $<$ 6\% of expansion, which is expected due to the cut-off as the outer parts are loosely bound and few particles could in principle escape.   
\begin{figure*}
\subfigure{\includegraphics[width=0.42\textwidth]{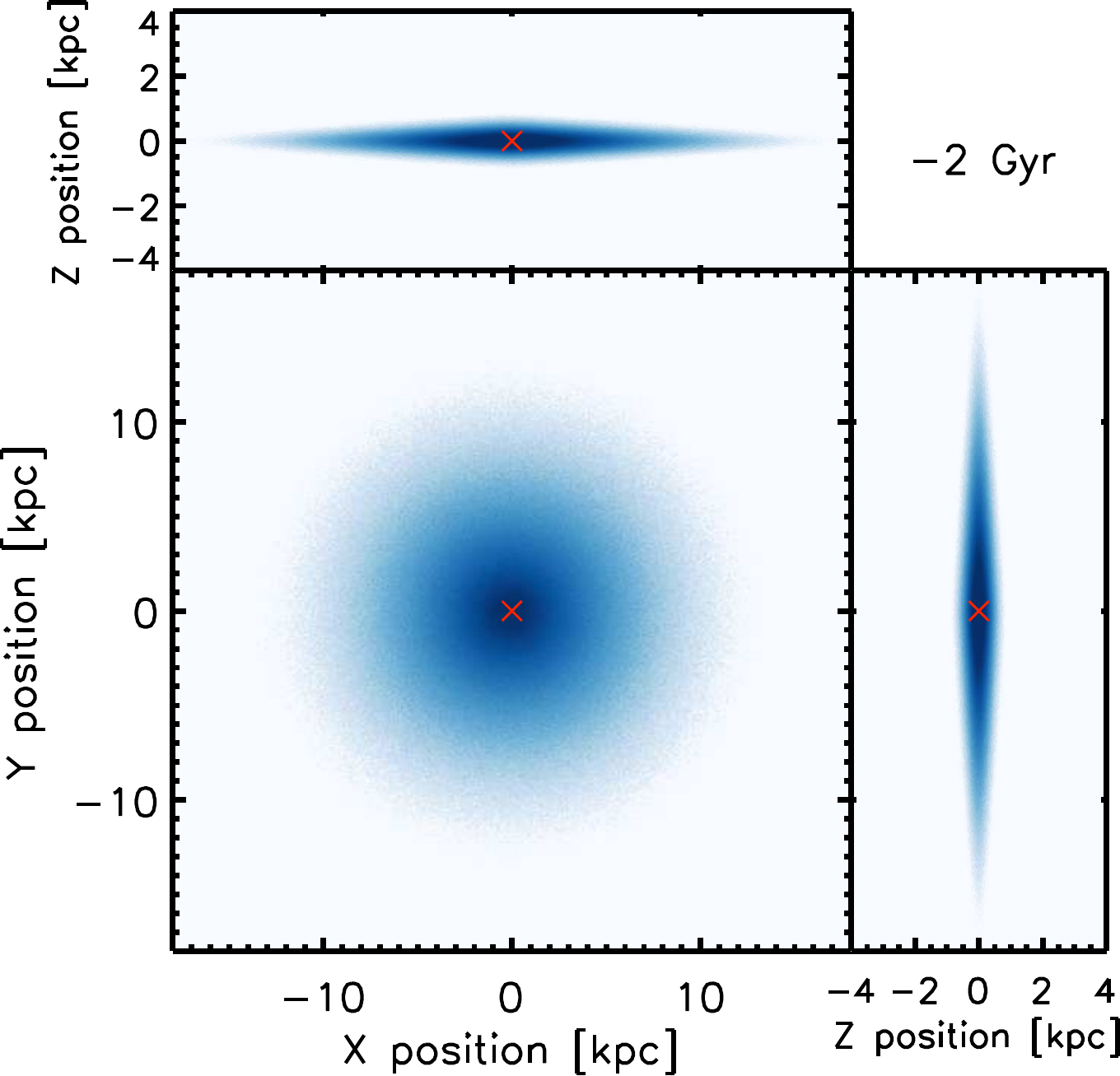}}
\subfigure{\includegraphics[width=0.51\textwidth]{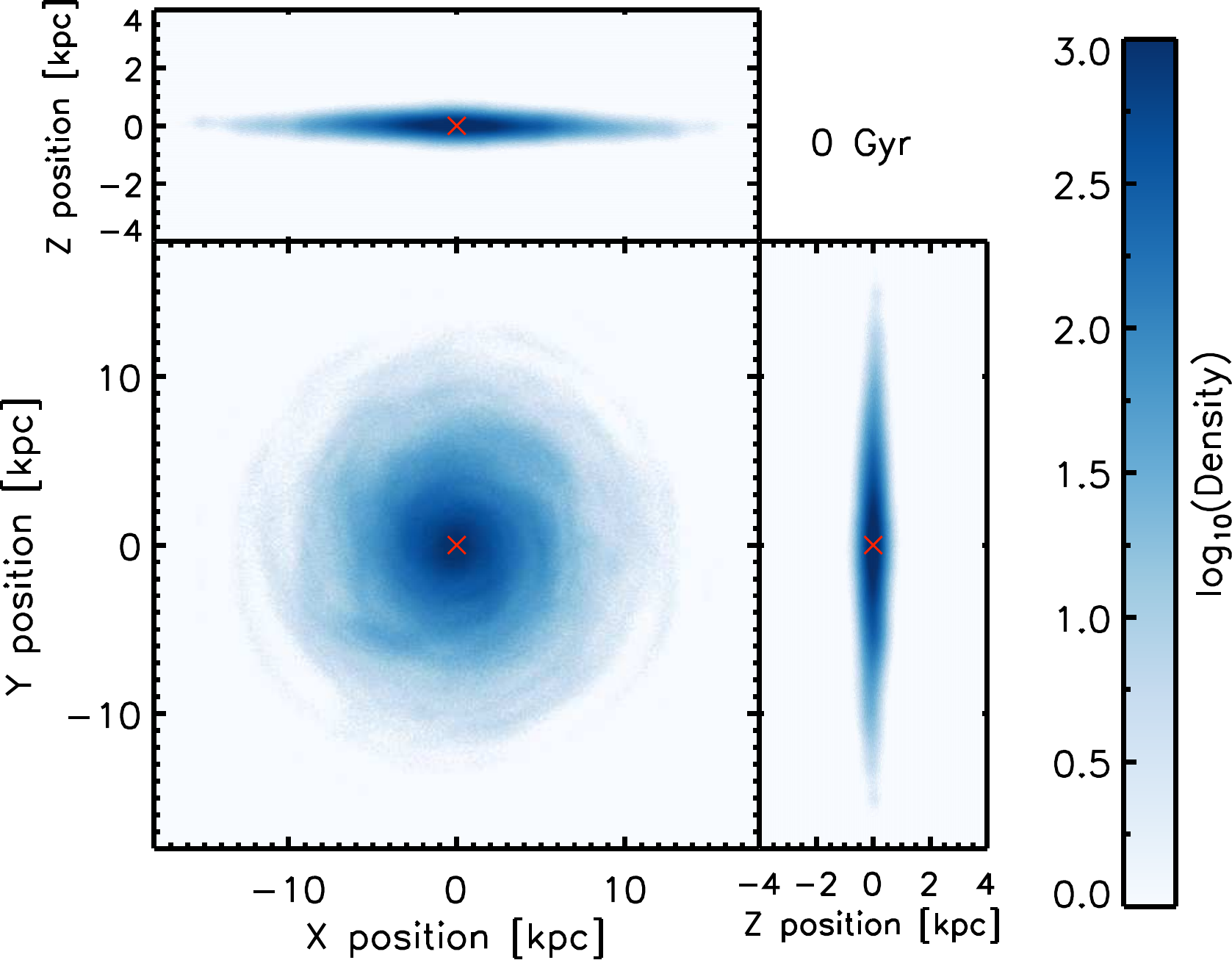}}
\subfigure{\includegraphics[width=0.42\textwidth]{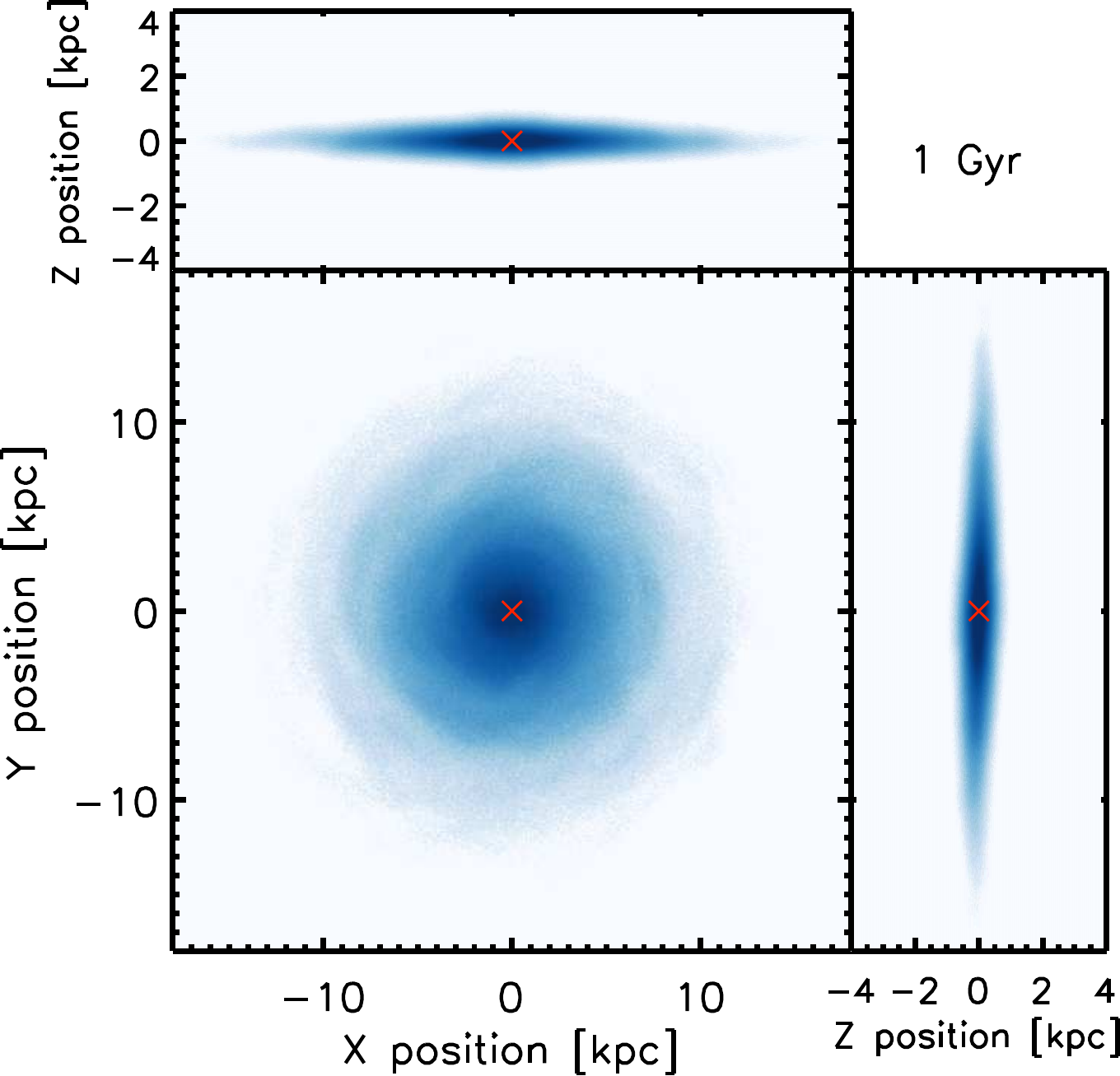}}
\subfigure{\includegraphics[width=0.51\textwidth]{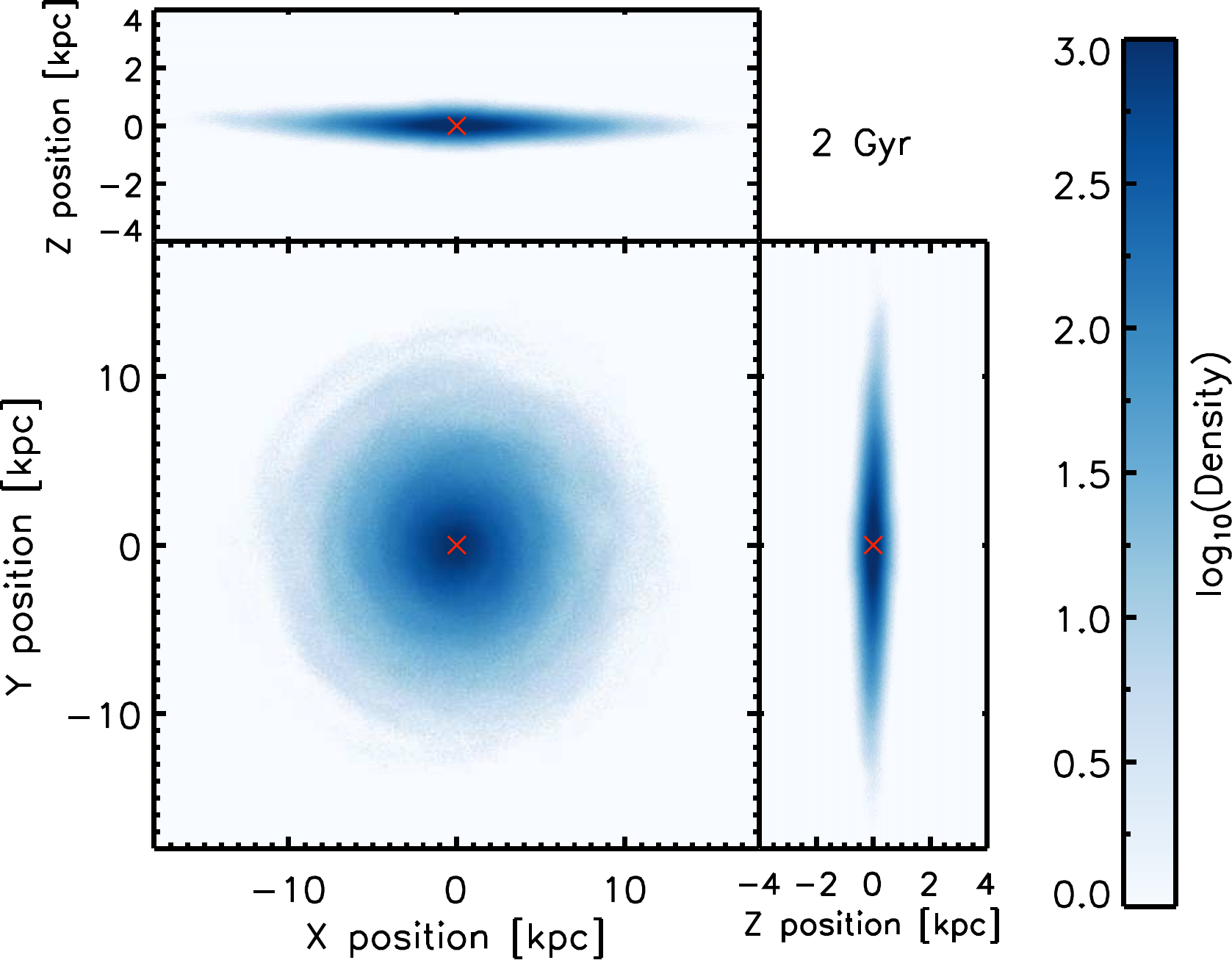}}
\caption{The number density plot for the isolated system in x-y, x-z and y-z view. \textit{From top left:} Aq-B2 simulation at initial time (-2 Gyr), 0, 1 and 2 Gyr of evolution. The red cross represents the centre of the mass of the bulge component.}
\label{fig:dens}
\end{figure*}

The number density plot of the disc from the Aq-B2 simulation is represented in Fig.\,\ref{fig:dens} where starting at top-left we observe the system at -2, 0, 1 and 2 Gyr. After 4 Gyr of evolution the disc appears in steady state and we do not observe any strong instabilities within the disc. Also the red crosses correspond to centre of mass of the bulge which coincides with the disc centre of mass.

\begin{figure*}
\includegraphics[width=\linewidth]{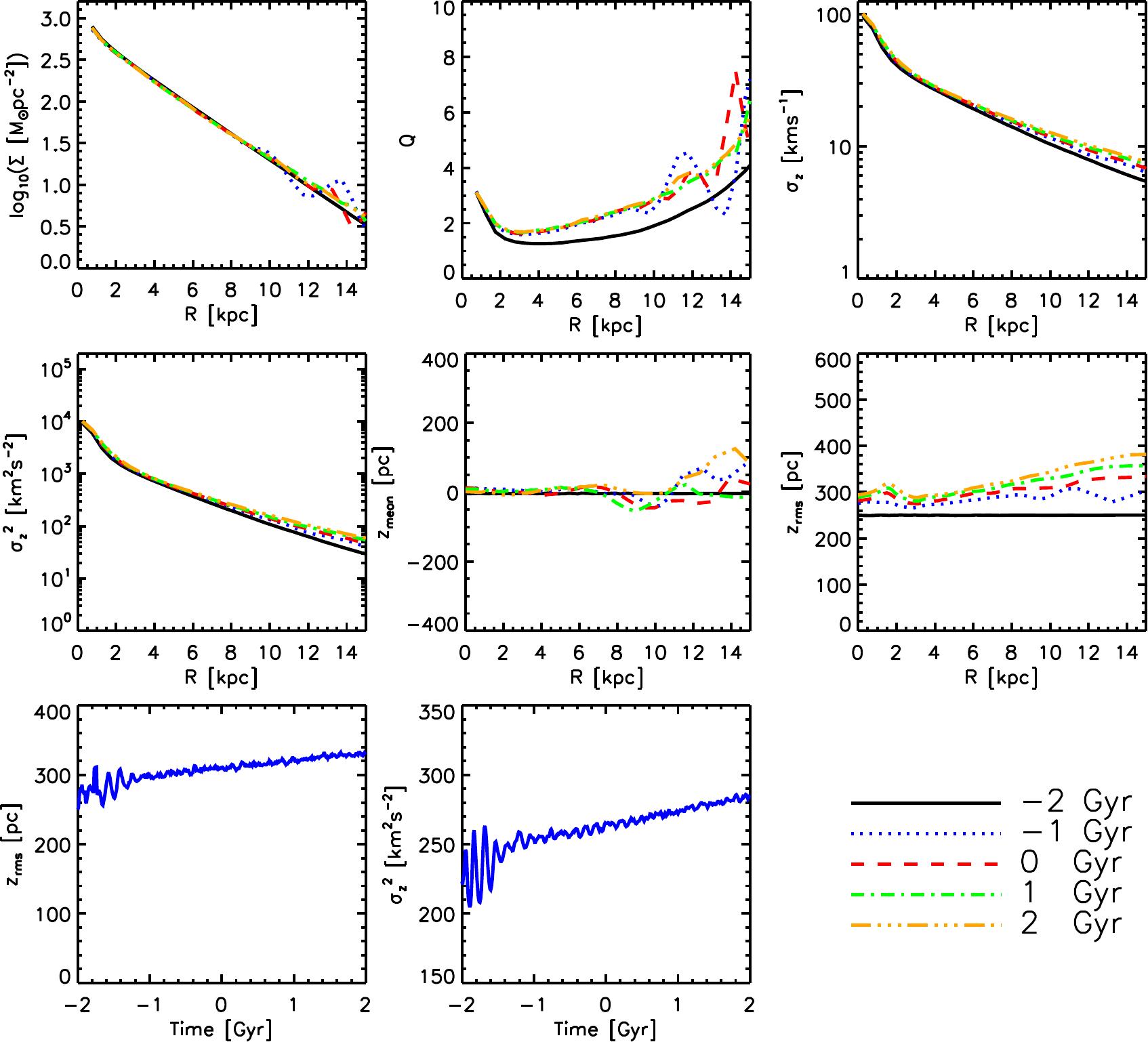}
\caption{The properties of the disc in the isolated Aq-B2 simulation. From the top left as function of radial distance from the centre of disc: Surface density of the disc $\Sigma$, Toomre parameter \textit{Q}, vertical velocity dispersion $\sigma_{z}$, vertical $\sigma_{z}^{2}$, mean disc height $z_\mathrm{mean}$, root-mean-square thickness $z_\mathrm{rms}$. The lines represent the results at -2, -1, 0, 1 and 2 Gyr. The two bottom figures show the $z_\mathrm{rms}$ and $\sigma_{z}^{2}$ in the solar neighbourhood, 7.5 $<$ R $<$ 8.5, as function of time.}
\label{fig:cont-multi}
\end{figure*}

Fig.\,\ref{fig:cont-multi} shows a few quantities describing the dynamics of the isolated system Aq-B2 -- a fair representative of our simulation sample. In the top two rows different line colours correspond to the system at -2, -1, 0, 1 and 2 Gyr of evolution as function of radial distance across the disc. Starting from the top left, the surface density $\Sigma$ of the disc is represented and stays unchanged throughout 4 Gyr of evolution in isolation. We observe some fluctuations at the outer disc, R $>$ 10\,kpc, in the initial 1 Gyr which dampens afterwards. The next plot shows the Toomre parameter \textit{Q}, a measure of the disc stability and is calculated as
\begin{equation}
Q=\frac{\sigma_{R}\kappa}{3.36G\Sigma},
\label{eq:toomre}
\end{equation}
where $\sigma_{R}$ and $\kappa$ represent the radial velocity dispersion and the epicyclic frequency of the disc particles. Initially the disc has a \textit{Q} value of 1.5 at the solar neighbourhood increasing to $\sim$ 2.5 after few hundred Myr. The initial increase of $\sigma_{R}$ is still one of the caveats of the GalIC code. The vertical velocity dispersion $\sigma_{z}$ shows an increase of 45\% at R\,=\,15\,kpc after 4 Gyr while for the solar neighbourhood we observe an increase of $<$ 12\% only. Accordingly, $z_\textrm{mean}$ increases by maximum of 25\,pc for the inner 8\,kpc while the outer part experiences an increase of 100\,pc after 4 Gyr. The next plot shows the time evolution of $z_\textrm{rms}$ in the solar neighbourhood. The increase of the disc thickness after 3 Gyr is $\sim$ 15 \% which reaches 320\,pc while in the case of the $\sigma_{z}^{2}$ in the solar neighbourhood a vertical heating of 40\,km$^{2}$\,s$^{-2}$ -- 16 \% -- is seen in the last 3 Gyr. 
\subsection{Full simulations}
After obtaining an isolated system which satisfies the equilibrium criteria, we add the subhalo candidates using their extracted initial positions and velocities at t=0\,Gyr. Since there is no massive satellite with $M_\textrm{tid} > 5 \times 10^{8} M_{\odot}$ closer than 20\,kpc (see Fig. \ref{fig:orbit-all}), we don't expect an unphysical reaction of the disc by the sudden implementation of the satellites. Each of the seven systems were simulated for 2 Gyr.

\section{Results}
\label{sec:results}
In this section we discuss the results obtained for the analysis of satellites' impact on the disc. We focus on the vertical heating of the galactic disc due to the infalling satellites, expressed in terms of thickening the disc and increasing the vertical velocity dispersion of disc particles. In the first part we discuss the heating in the solar neighbourhood, 7.5\,kpc $<$ R $<$ 8.5\,kpc, for the duration of 2 Gyr while in the second part we analyse the impact of satellites across the whole disc radial range. In the last part, we argue about the relevance of the orientation of satellites' distribution to the vertical heating measurements for two cases of Aq-D2 rotated along the $x$ and $y$ axes by 90 degrees. 
  
\subsection{Solar neighbourhood}
\begin{figure}
\includegraphics[width=\linewidth]{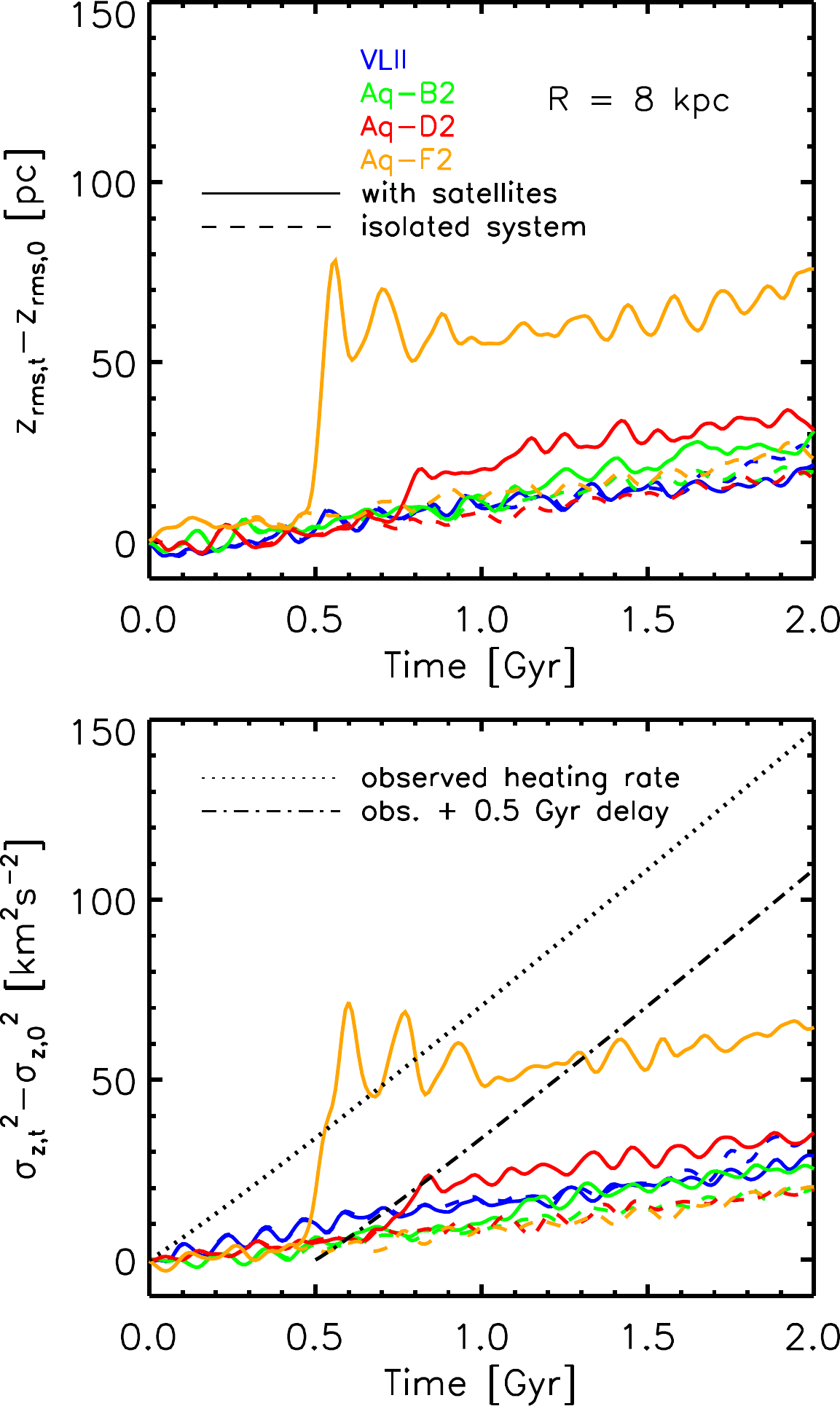}
\caption{\textit{Top:} Time evolution of the root-mean-square of the disc thickness subtracted by the initial value for systems with satellites (solid) and isolated (dashed) for VL\rom{2} (blue), Aq-B2 (green), Aq-D2 (red) and Aq-F2 (orange) simulations in the solar neighbourhood. \textit{Bottom:} The time evolution of the difference of the disc vertical heating for VL\rom{2} (blue), Aq-B2 (green), Aq-D2 (red) and Aq-F2 (orange) simulations. The dotted and dot-dashed lines represent the observed vertical heating rate in the solar neighbourhood and the observed value with 0.5 Gyr delay, respectively.}
\label{fig:zstd-sigz2-time}
\end{figure}

In this study, with advantage of our high resolution analysis of the Milky Way disc, we are able to resolve the structure of the disc at 40\,pc scale together with the large number of disc particles. The next results focus on the time evolution of the $z_\textrm{rms}$ and vertical velocity dispersion in the solar neighbourhood for all the particles lying in the radial range 7.5 - 8.5\,kpc from the disc centre. The top panel of Fig.\,\ref{fig:zstd-sigz2-time} shows the time evolution of the root-mean-square of the disc height subtracted by the initial values, $z_\textrm{rms,t}-z_\textrm{rms,0}$ for systems with satellites (solid) and isolated (dashed) at the solar neighbourhood for VL\rom{2} (blue), Aq-B2 (green), Aq-D2 (red) and Aq-F2 (orange). During the initial 0.4 Gyr the thickening is minimal since no massive subhalo with $M_\textrm{tid}$ $\geq$ 10$^{9}$ $M_{\odot}$ crosses the disc within this period (Fig.\,\ref{fig:orbit-all}). The reason for this lack of interaction in the first 0.4\,Gyr may be the result of initial conditions at $z$=0, since we did not take into account the tidal mass loss which already happened for satellites in the inner region of the host halo\,\citep{nagai}. In order to take this into account; we will compare the dynamical evolution in the simulations in addition to a delayed heating by 0.5\,Gyr. Aq-F2 reaches a maximum thickness difference of 75\,pc ($\sim$ twice as much as in the isolated case at 2 Gyr); while in the remaining simulations the thickness increases by 30\,pc at most.

As mentioned in section\,\ref{sec:int}, the primary aim of this study is to understand the contribution of realistic cosmological satellites to the heating of the galactic disc. This process can be best quantified using the squared vertical dispersion $\sigma_{z}^{2}$ of the disc particles. The most reliable measurement of the vertical heating of the Milky Way disc has been obtained for the solar neighbourhood (e.g. \citealt{holmberg07,holmberg09}) with the value of $d(\sigma_{z}^{2})/dt$\,=\,72\,km$^{2}$\,s$^{-2}$\,Gyr$^{-1}$ following a power law with exponent $2\gamma$\,=\,1.06 for $\sigma_{z}^{2}$ which means a constant heating rate. The time evolution of the quantity $\sigma_{z,t}^{2}$ - $\sigma_{z,0}^{2}$ (the difference between the system at time \textit{t} and the initial time when the satellites are added) for systems with satellites and isolated in the solar neighbourhood is presented in the bottom panel of Fig.\,\ref{fig:zstd-sigz2-time} with $\sigma_{z,0}^{2}$\,=\,260\,km$^{2}$\,s$^{-2}$. As with the evolution of $z_\textrm{rms}$ we observe a more significant heating in the case of Aq-F2 -- 25\% at 2 Gyr compared to initial -- while only very small fluctuations are seen for the remaining simulations. The self-heating of the disc is very small, on scale of $\sim$ 10 km$^{2}$\,s$^{-2}$\,Gyr$^{-1}$. The dotted and dot-dashed lines, respectively, represent the observed vertical heating rate in the solar neighbourhood and the observed value allowing for 0.5 Gyr delay which accounts for possible initial underestimation of heating due to mass loss of subhaloes already located in the inner halo. The jumps observed in the velocity dispersion due to massive mergers are seen in other simulations where such mergers cause visible signatures in the velocity dispersion profiles\,\citep*{martig}.

Aq-D2 has twice as many subhaloes with $M_\textrm{tid}$ $\geq$ 10$^{9}$ $M_{\odot}$ compared to Aq-B2 which, in theory, translates into a larger impact. We also observe twice as much heating compared to the isolated cases between the two simulations in the solar neighbourhood.

\begin{figure}
\includegraphics[width=\linewidth]{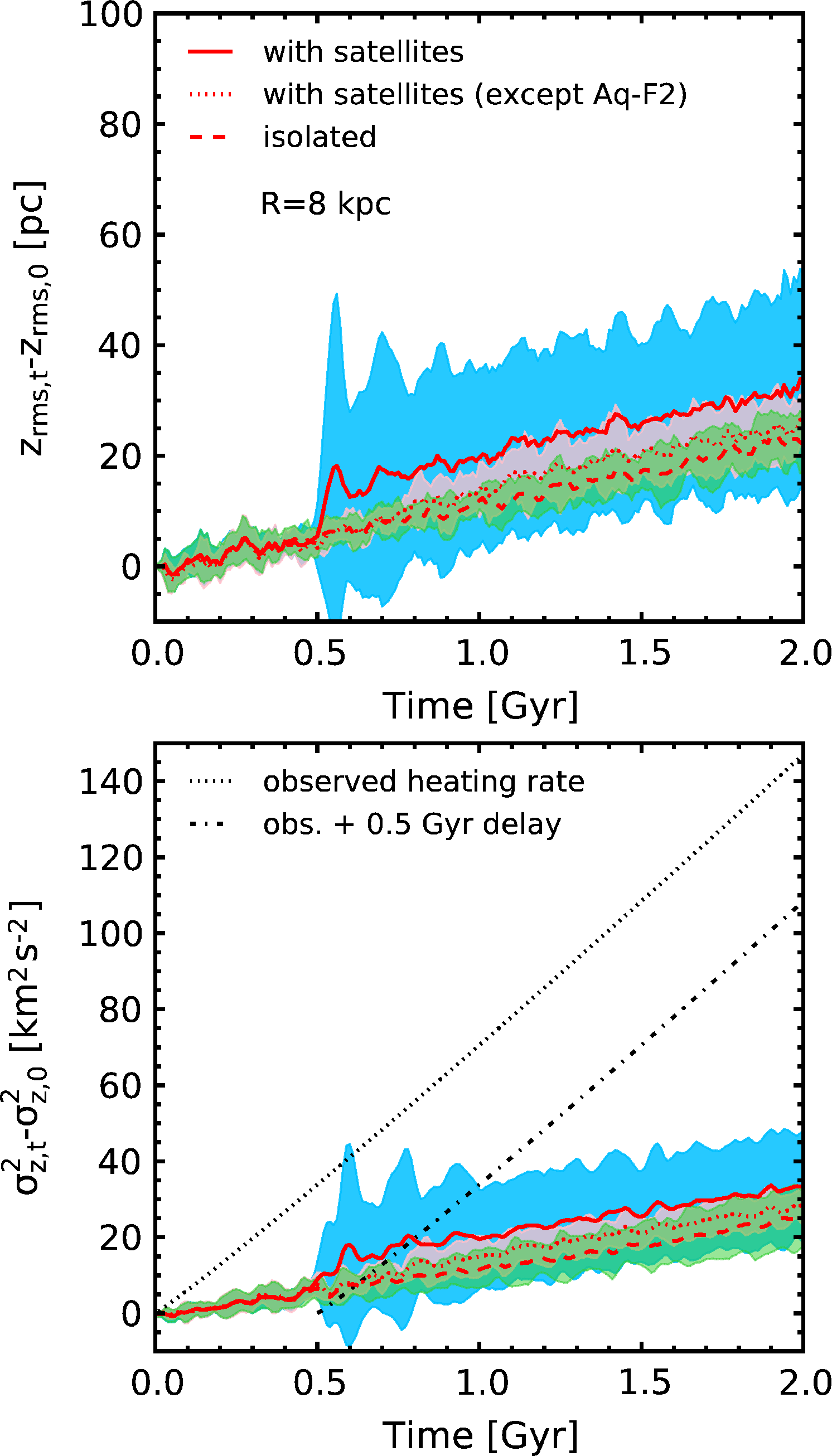}
\caption{The mean difference of vertical heating between t\,=\,2 and t\,=\,0 Gyr together with their 1$\sigma$ statistical dispersion region of all 7 systems with satellites (solid, blue), with satellites except Aq-F2 (dotted, pink) and isolated (dashed, green). The dotted and dot-dashed lines represent the observed vertical heating rate in the solar neighbourhood and the observed value with 0.5 Gyr delay.}
\label{fig:zstd-sigz2-time-disp}
\end{figure}

The mean values of $z_\textrm{rms,t}-z_\textrm{rms,0}$ for all 7 simulations with satellites (solid), with satellites except Aq-F2 (dotted) and isolated system (dashed) are shown in the top panel of Fig.\,\ref{fig:zstd-sigz2-time-disp} with their subsequent 1$\sigma$ statistical dispersion around the mean. The blue, pink and green colours correspond to all the simulations with satellites, with satellites except Aq-F2, and for the isolated system, respectively. We observe a very shallow increase of the thickness with time in all systems after the initial 0.4 Gyr. The mean value of all the systems with satellites experiences an increase of 35\,pc compared to the initial state and only $<$ 10\,pc -- $\sim$ 40\% -- compared to the isolated system. If we exclude the impact from Aq-F2, the thickening in the solar neighbourhood from satellites is almost indistinguishable from the isolated case. 

According to the bottom panel of Fig.\,\ref{fig:zstd-sigz2-time-disp} there exists a slow increasing trend of the mean value of $\sigma_{z,t}^{2}$ -- $\sigma_{z,0}^{2}$ for all systems. The heating rate for the initial 1 Gyr are 20, 15 and 10\,km$^{2}$\,s$^{-2}$\,Gyr$^{-1}$ for systems with satellites, with satellites except Aq-F2, and the isolated system, respectively, which reduce to 15\,km$^{2}$\,s$^{-2}$\,Gyr$^{-1}$ for all the systems later. This means the rate at which the vertical dispersion increases with time in the solar neighbourhood decreases slightly after 1 Gyr. Not taking into account Aq-F2, the difference between the mean heating due to satellites and the isolated systems is very small and only 5\,km$^{2}$\,s$^{-2}$ at 2 Gyr. The 1$\sigma$ dispersion region from all the simulations (blue) reaches 50\,km$^{2}$\,s$^{-2}$ at the final snapshot which is 35\% of the observed value of 144\,km$^{2}$\,s$^{-2}$. The additional heating for subhaloes below 10$^{10} M_{\odot}$ is of the order of $\sim$ 10-15\%. The dotted and dot-dashed lines represent the observed vertical heating rate in the solar neighbourhood and the observed value with 0.5 Gyr delay, respectively. The Aq-F2 simulation suggests in order to reach the observed heating rate, that a few encounters per Gyr of LMC type satellites with pericentres below 20 kpc are required. Our result fails by more than 3$\sigma$ to reproduce the observed heating rate. These results are consistent with the recent observational work by\,\citet{ruchti}. Using the \textit{Gaia}-ESO spectroscopic survey, they found no evidence for a significant dark matter disc in the MW and that the Galactic disc has had a rather quiescent merger history for the past $\sim$ 9 Gyr.  

\subsection{The impact across the disc}
In addition to the solar neighbourhood, the global vertical heating across the whole disc was also investigated and is presented in this section. 

\begin{figure}
\includegraphics[width=\linewidth]{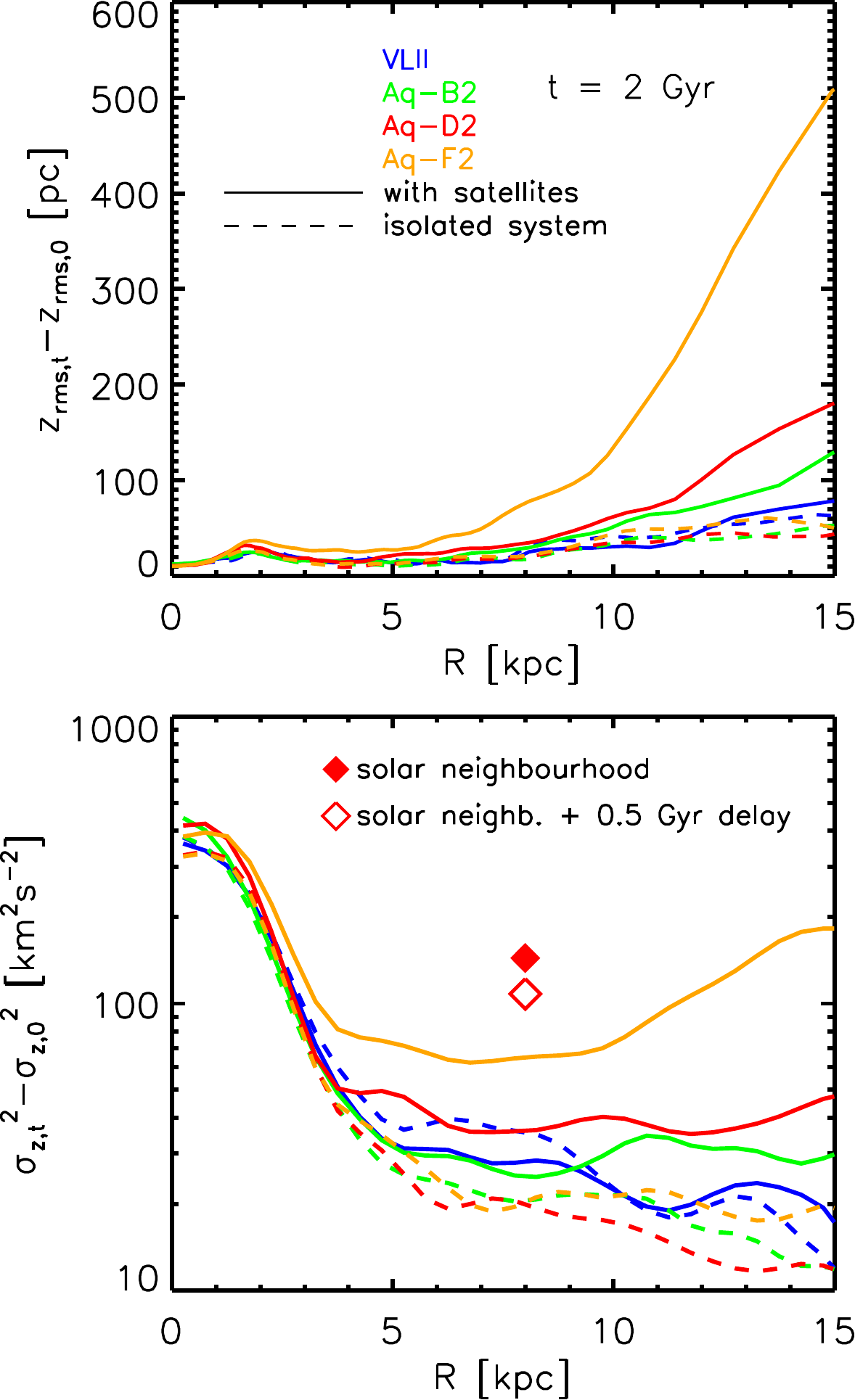}
\caption{\textit{Top:} The difference of root-mean-square of the disc height $z_\textrm{rms}$ at 2 and 0 Gyr for systems with satellites (solid) and isolated (dashed) for VL\rom{2} (blue), Aq-B2 (green), Aq-D2 (red) and Aq-F2 (orange). \textit{Bottom:} The difference of the disc vertical heating between t\,=\,2 and 0 Gyr for VL\rom{2} (blue), Aq-B2 (green), Aq-D2 (red) and Aq-F2 (orange). filled and open \textit{diamonds} represent the observed value of vertical heating in the solar neighbourhood and the observed value with 0.5 Gyr delay.}
\label{fig:zstd-sigz2}
\end{figure}

The root-mean-square of the disc height at 2 Gyr subtracted by the initial values, $z_\textrm{rms,t}-z_\textrm{rms,0}$, as a function of radial distance from the disc centre for systems with satellites (solid) and isolated (dashed) for VL\rom{2} (blue), Aq-B2 (green), Aq-D2 (red) and Aq-F2 (orange) simulations are shown in the top panel of Fig.\,\ref{fig:zstd-sigz2}; this quantity is calculated in radial bins of equal particle numbers. The vertical heating ($\sigma_{z,t}^{2}$-$\sigma_{z,0}^{2}$) of the disc is shown in the bottom panel of Fig.\,\ref{fig:zstd-sigz2}. The solid and dashed lines illustrate the systems with satellites and the isolated control systems; the results correspond to the time of 2 Gyr which is the final snapshot of our simulations. The deviation between the control system and the system with satellites is mainly visible in the outer disc. In Aq-F2 the flare starts at the solar neighbourhood and reaches a thickening of 550\,pc at R\,=\,15\,kpc. For the VL\rom{2} simulation we observe that the thickening of the isolated case is slightly larger than the case with satellites at some radii, however we still observe a positive increase of thickness compared to 0 Gyr time. As seen in previous studies\,\citep{ardi} we observe a small flaring in the outer parts of the disc which is amplified with the presence of the satellites; the particles which reside in the outer regions are less strongly bound to the disc and require less energy to be displaced. Aq-A2, Aq-C2 and Aq-E2 have the least thickening due to infalling satellites.

The bottom panel of Fig.\,\ref{fig:zstd-sigz2} illustrates the quantity $\sigma_{z,t}^{2}$ - $\sigma_{z,0}^{2}$. Aq-F2 dominates the heating -- reaching $\sim$ 200\,km$^{2}$\,s$^{-2}$ in the outer parts. The peak at R$<$4\,kpc corresponds to a very small increase in velocity dispersion (from 100\,km\,s$^{-1}$ to 102\,km\,s$^{-1}$ at the centre). The observed vertical heating in the solar neighbourhood after 2 Gyr\,\citep{holmberg09} is represented as a filled diamond while if we allow for 0.5 Gyr of delay we obtain the heating represented by the empty diamond in Fig.\,\ref{fig:zstd-sigz2} with values of 144 and 108\,km$^{2}$\,s$^{-2}$, where 144 corresponds to twice the vertical heating rate of 72\,km$^{2}$\,s$^{-2}$\,Gyr$^{-1}$. The $\sigma_{z,0}^{2}$ at R\,=\,15\,kpc has a value of 40\,km$^{2}$\,s$^{-2}$ and therefore the maximum heating in the outer parts reaches $\sim$ 500\% compared to the initial value -- excluding Aq-F2 we only have $<$ 130\%.

Similar to the behaviour seen in the solar neighbourhood, Aq-B2 and Aq-D2 possess comparable heating rates with 20\,km$^{2}$\,s$^{-2}$ extra heating for Aq-D2. Aq-D2 has half of its pericentric passages during the initial 1.1 Gyr while the remaining passages occur during the final 0.1 Gyr. Hence it appears that the heating due to subhaloes with 10$^{9}$ $<$ $M_\textrm{tid}$ $M_{\odot}$ $<$ 10$^{10}$ $M_{\odot}$ contributes only to $\sim$ 40\,km$^{2}$\,s$^{-2}$ of heating for R $>$ 4\,kpc. 

\begin{figure}
\includegraphics[width=\linewidth]{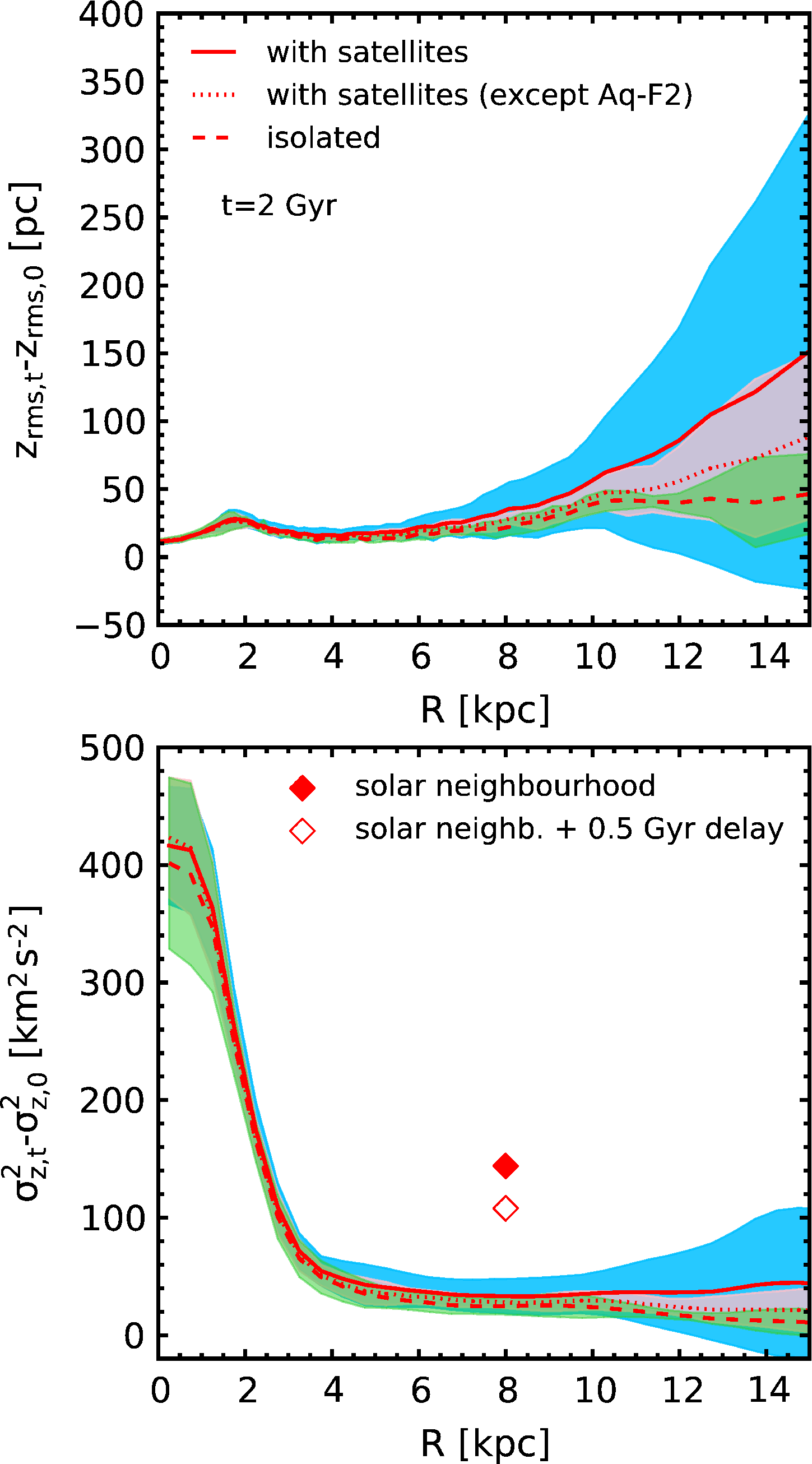}
\caption{\textit{Top:} The mean root-mean-square of the disc height for all 7 systems with satellites (solid), with satellites except Aq-F2 (dotted) and all 7 isolated system (dashed) at 2 Gyr. \textit{Bottom:} The mean difference of the disc vertical heating between t\,=\,2 and 0 Gyr for similar systems as in the top panel. The shaded regions correspond to the 1$\sigma$ dispersion around the mean value for all 7 systems with satellites (blue), with satellites except Aq-F2 (pink) and isolated systems (green). Filled and open \textit{diamonds} represent the observed value of vertical heating in the solar neighbourhood and the observed value with 0.5 Gyr delay.}
\label{fig:zstd-sigz2-disp}
\end{figure}

The top panel of Fig.\,\ref{fig:zstd-sigz2-disp} demonstrates the mean $\sigma_{z,t}^{2}$ - $\sigma_{z,0}^{2}$ value of all seven simulations with satellites (solid), with satellites except Aq-F2 (dotted) and the isolated system (dashed), across the radial range at t\,=\,2 Gyr. The shaded regions present the 1$\sigma$ statistical dispersion in each bin. The impact of Aq-F2 is seen as the difference of the dotted and solid lines which increases as we move towards outer regions. The mean disc height of all simulations with satellites increases by 150\,pc after 2 Gyr at 15\,kpc and including the dispersion this value reaches 340\,pc; while the isolated systems only have an increase of 50\,pc at this radius, corresponding to self thickening of about 14\%. Aq-F2 increases the mean value by an additional 70\,pc in the outer 15\,kpc regime.

The mean value of the vertical heating between 0 and 2 Gyr for all systems across the disc is shown in the bottom panel of Fig.\,\ref{fig:zstd-sigz2-disp}. The shaded areas correspond to the 1$\sigma$ dispersion around the mean for the subsequent systems -- blue, pink and green. For all the systems with satellites, the mean value lies around 40\,km$^{2}$\,s$^{-2}$ across the disc at R$>$4\,kpc, while the 
simulations excluding Aq-F2 possess $<$ 20\,km$^{2}$\,s$^{-2}$ in the outer parts. The solar neighbourhood heating at R=8\,kpc of 144\,km$^{2}$\,s$^{-2}$ and the value 108\,km$^{2}$\,s$^{-2}$ taking into account 0.5 Gyr delay are represented as filled and open diamonds. The 1$\sigma$ dispersion in the solar neighbourhood reaches only 50\,km$^{2}$\,s$^{-2}$. We observe a specific heating almost independent of galactocentric distances for R $\ge$ 4\,kpc. More energy is deposited to the inner parts; although, this region is not strongly perturbed due to both higher surface density and velocity dispersion. However, the outer parts with lower surface density experiences stronger relative heating together with flaring. 

\subsection{Effect of satellites infall orientation}
In this section we present our results from the analysis of the impact of the orientation of satellites distribution relative to the disc on the measured vertical disc heating. For this purpose we only use the Aq-D2 simulation, which has the second most number of satellites candidates (23) above 10$^{8}$ $M_{\odot}$ and the most candidates with $M_\textrm{tid}$ $\geq$ 10$^{9}$ $M_{\odot}$. This simulation is believed to be representative of our simulation suites and shows a typical level of anisotropy (Fig.\,\ref{fig:incl_hist}). We decided to consider two additional cases of different orientation that consist of rotating the satellites' frame by 90 degrees along the $x$ axis, hereafter case B, and also performing the similar rotation along the $y$ axis, case C. The original orientation is regarded as case A. These rotations correspond to changing the disc original axis of rotation, $z$, to rotation about the $y$ (case B) and $x$ axis (case C). 
\begin{figure}
\includegraphics[width=\linewidth]{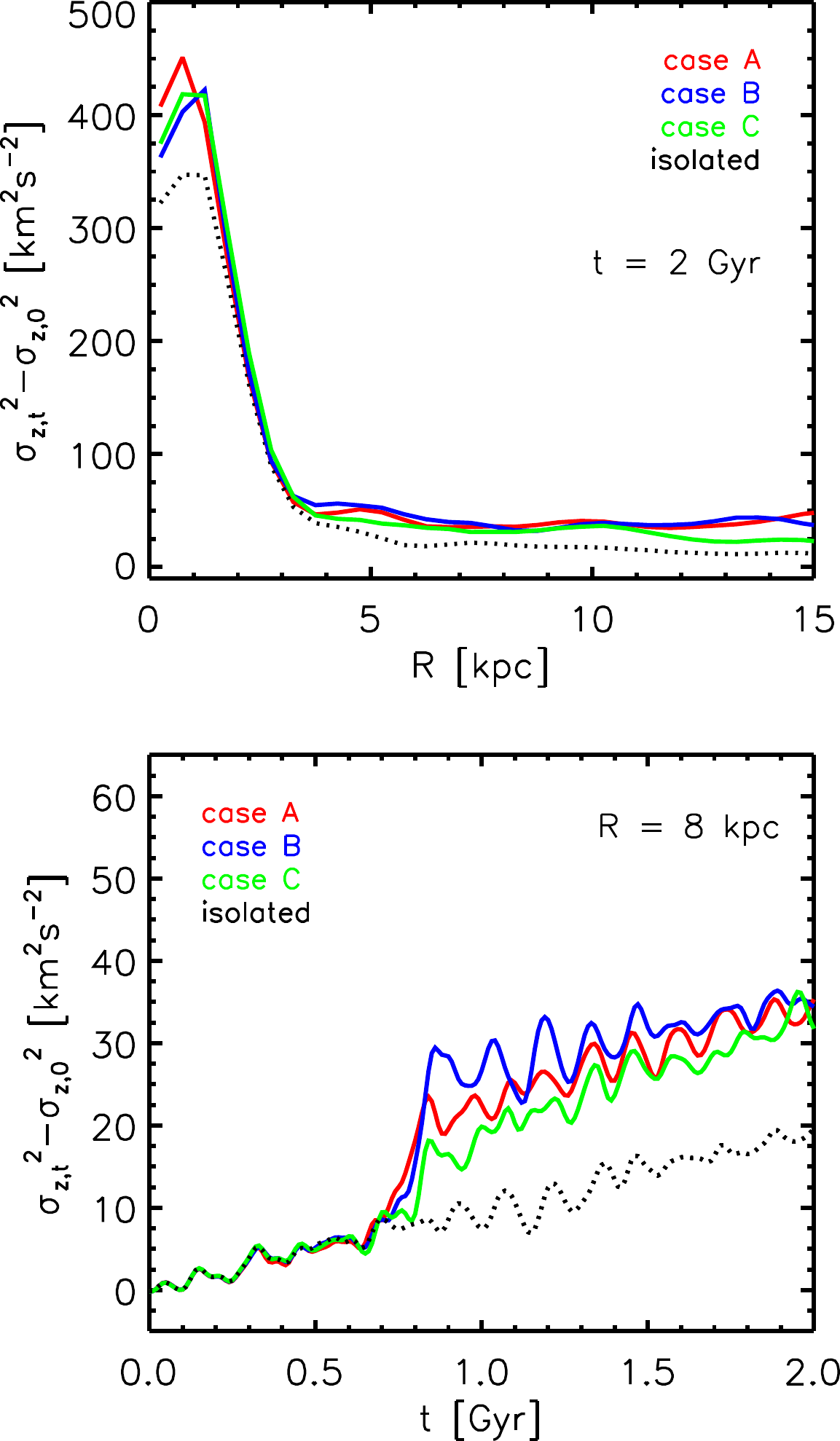}
\caption{\textit{Top:} The difference of the disc vertical heating for original Aq-D2 case A (red), case B (blue), case C (green) and isolated (dotted) systems across the disc at the final snapshot, t\,=\,2 Gyr. \textit{Bottom:} the time evolution of the difference of the disc vertical heating for similar systems as in the top panel in the solar neighbourhood.}
\label{fig:sigz2-D2}
\end{figure} 

The top panel of Fig.\,\ref{fig:sigz2-D2} shows the difference of the disc vertical heating, $\sigma_{z,t}^{2}$ -- $\sigma_{z,0}^{2}$, for the original Aq-D2 case A (red), case B (blue), case C (green) and isolated (black) systems across the disc at the end of the simulation, t\,=\,2 Gyr. The heating across the disc is very similar. Only in case C the outer disc shows a reduced heating. The time evolution of the same quantity in the solar neighbourhood is shown in the bottom panel of Fig.\,\ref{fig:sigz2-D2}. The difference is minimal during the initial 0.4 Gyr when no massive subhalo crosses the disc. The time evolution differs by 10\,km$^{2}$\,s$^{-2}$ around 1 Gyr. However, the final heating rate is the same in all cases. Hence the impact of orientation plays a minor role in the vertical heating of the galactic disc.

\section{Summary}
\label{sec:summary}
In this study we have attempted at understanding the significance of the impact of satellites on the dynamics of the Milky Way disc using realistic initial conditions for the DM substructures from the zoom-in cosmological simulations of Milky Way-like dark matter haloes. Such impacts are analysed in the form of vertical heating of the galactic disc in the presence of satellite galaxies. This work benefits from four major assets: 1. N-body simulations which reach resolutions of 40 parsecs for the disc in the vertical direction, 2. initial conditions for the multi-component Milky Way galaxy which resides in much better equilibrium and steady state in the isolated case compared to previous works, 3. realistic initial conditions for DM substructures and 4. statistical set of initial conditions performed for the first time. 

To have a statistically robust understanding of the satellites infall, we used a distribution of subhaloes from seven cosmological simulations including six Aquarius simulations (at level 2 resolution) together with the Via Lactea \rom{2}. We used the $z$\,=\,0 snapshots containing a few hundred thousands DM subhaloes. The masses, velocities and positions of all the subhaloes were normalised using the ratio between the host DM halo masses ($M_{200}$) of the corresponding simulations and Aq-D2. The subhaloes at the low mass end of the distribution only consist of a few hundred particles, therefore we decided to consider only objects with $M_\textrm{tid}$ $\geq$ 10$^{6}$ $M_{\odot}$ for all our analysis. The fraction of total DM mass residing within substructures, $f_\textrm{sub}$, is a measure of the clumpiness and ranges from 5\% for VL\rom{2} to 14\% within r$_{50}$ of the host halo. Such a range is an indication for the cosmic scatter between the DM mass distribution of different simulations. In all simulations this fraction drops by three orders of magnitude as we move towards the inner parts of the main haloes. The distribution of subhaloes within the zoom-in cosmological simulations possesses differences between various simulation suites. VL\rom{2} has the least isotropic distribution of DM substructures compared to Aquarius. The subhaloes appear to be preferentially located around the $z$-axis in VL\rom{2} while we also observe an anisotropic distribution in the $x$-$y$ plane.  

The orbits of these subhaloes were integrated in the presence of their background host halo potential to identify the subhaloes which come closer than 25\,kpc to the centre. The percentage fraction of these crossed subhaloes ranges from 6.9\% to 14.5\%. The cumulative subhalo mass distribution and the maximum circular velocity distributions of the crossed subhaloes show a similar slope as the total population, in the case of all 7 simulations, implying the crossed sample is a good representative of the total. The crossed candidates located at $r$ $>$ 100\,kpc contribute to $<$ 20\% of total number of subhaloes at the corresponding radii, while more than 50\% of the \textit{total} crossed objects originate from this outer region. There exists an increasing trend for the concentration of subhaloes, c$_\textrm{tid}$\,=\,$r_\textrm{tid}$/$r_\textrm{s}$, with increasing distance. Crossed candidates located at $r$ $>$ 100\,kpc have lower c$_\textrm{tid}$ than the total population. Considering the large contribution of objects at this distance to the crossed sample together with higher mean concentration of the total case, crossed candidates have a shift towards lower c$_\textrm{tid}$ across the whole mass range by a factor of 1.35. 

The impact strength of these subhaloes is expected to scale $\propto$ $M_\textrm{tid}^{2}$, hence we chose only the crossed subhaloes with $M_\textrm{tid}$ $\geq$ 10$^{8}$ $M_{\odot}$ to use in our N-body simulations. Aq-F2 has the most number of candidates satisfying this criteria with 24 subhaloes, while Aq-E2 has the least with only 12. Four subhaloes with $M_\textrm{tid}$ $\geq$ 10$^{9}$ $M_{\odot}$ reside in Aq-D2 and Aq-F2 is the only simulation that has one candidate with $M_\textrm{tid}$ $\geq$ 10$^{10}$ $M_{\odot}$. The initial conditions for our Milky Way galaxies were generated using the GalIC code\,\citep{yurin} with a thin exponential disc, a Hernquist bulge and a matching Hernquist DM halo for the original NFW host haloes. The parallelised N-body particle-mesh code SUPERBOX-10 was used for performing the simulations with three nested grids of different resolutions per galaxy. In the case of the disc we used 10 million particles in order to suppress any unwanted perturbations caused by low number statistics in the disc. The isolated MW systems were evolved for 2 Gyr so that the components can adapt to their environment and also reach an equilibrium state. The subhalo candidates with $M_\textrm{tid}$ $\geq$ 10$^{8}$ $M_{\odot}$ for each simulation were inserted according to their initial positions and velocities extracted from the cosmological simulations. These systems were evolved for 2 Gyr while the isolated cases were also simulated in parallel as control group. The vertical heating of the galactic disc was determined by measuring the root-mean-square of the disc thickness $z_\textrm{rms}$ and the vertical velocity dispersion squared $\sigma_{z}^{2}$ in the solar neighbourhood and across the disc. We observe a flaring in the outer parts of the disc where the mean value at R\,=\,15\,kpc after 2 Gyr for increase of the thickness due to satellites is about 100\,pc compared to the isolated case, including all seven simulations. However, if we exclude Aq-F2 the increase is only 30\,pc. If we take into account the 1$\sigma$ statistical dispersion around the mean, the thickness of the disc can reach 700\,pc at the outer parts at 2 Gyr. It is important to note that it is easier to increase the vertical thickness of the particle distribution in the outer regions of the disc than in the inner parts since these particles are less strongly bounded to the disc. The mean value of $\sigma_{z,t}^{2}$ - $\sigma_{z,0}^{2}$ at the end of our simulation drops strongly in the inner 4\,kpc of the disc while becoming constant as we move outwards. This heating is 40\,km$^{2}$\,s$^{-2}$ for all the simulations and $<$ 20\,km$^{2}$\,s$^{-2}$ excluding Aq-F2. This means that the specific energy input is homogeneously distributed over the disc in the radial range 4-15\,kpc. The isolated system also experiences self-heating of a similar value as the systems with satellites, neglecting Aq-F2. 

According to the work done by\,\citet{holmberg09}, the observed vertical heating in the solar neighbourhood after 2 Gyr is expected to be 144\,km$^{2}$\,s$^{-2}$. The time evolution of the heating in the solar neighbourhood was analysed and a slowly increasing trend for $\sigma_{z,t}^{2}$ - $\sigma_{z,0}^{2}$ was observed in all our  systems. For the first 0.4 Gyr a minimal heating was observed, since no massive satellite with $M_\textrm{tid}$ $\geq$ 5 $\times$ 10$^{8}$ $M_{\odot}$ crosses the disc within this period. However, the heating rate in the initial 1 Gyr is slightly higher than, in the final 1 Gyr. The observed value of 144\,km$^{2}$\,s$^{-2}$ lies more than 3$\sigma$ above the mean vertical heating taking into account all simulations. According to our analysis, the choice of the orientation of satellites distribution for the Aq-D2 case affects the measured vertical heating by less than 8\% after 2 Gyr in the solar neighbourhood, which is lower than the self-heating of the isolated system. The orientation plays only a minor role in heating the disc vertically. 

The measured heating from this analysis is sub-dominant and significantly smaller than in previous studies (e.g. \citealt{ardi,kazan09}). This mainly arises from the fact that our cosmological simulations do not include many subhaloes with $M_\mathrm{tid}$/$M_\mathrm{disc}$ $>$ 0.2, as was the case in other works, and are thought to heat the disc more significantly. Moreover, the DM clumpiness in Aquarius and VL\rom{2} is much smaller than that in earlier cosmological simulations, as they suffered from lower mass resolution. The insufficient heating in our simulations could also be subject to further reduction if we allow for the presence of gas in the disc\,\citep{moster}. Alternative mechanisms are more likely to dominate the disc heating. Heating by transient spiral arms or by disc growth via mass accretion are discussed in other works (e.g. \citealt*{villalobos2,martig}). The increase of disc surface density results in enhanced $\sigma_{z}$ with contribution towards heating, ranging from less than 5\% to 60\% for increasing surface density -- playing an important role in simulations which allow for disc growth. Coherent bending waves and buckling instabilities are other possible scenarios responsible for increasing the vertical dispersion of disc particles although \,\citet{toomre83} and\,\citet{sellwood} have shown such mechanisms are inefficient in terms of reproducing the observed vertical heating. Also\,\citet{lacey} and\,\citet*{ida} found that the scattering of disc stars due to GMCs is able to redirect the peculiar velocities of stars out of the plane however the inputted energy is ineffective in increasing the velocities. 

The results of this study show that the impact from the subhaloes on the vertical heating of the galactic disc is highly dependent on the high-mass end of the cosmological subhalo distribution -- $M_\textrm{tid}$ $\geq$ 10$^{9}$ $M_{\odot}$. The differences between different simulations can be thought as cosmic variance calculated as the statistical dispersion around the mean value from all simulations. If we exclude the Aq-F2 simulation as the only simulation with 1 subhalo of $M_\textrm{tid}$\,=\,5.9 $\times$ 10$^{10}$ $M_{\odot}$, the impact of satellites reaches only 3\,km$^{2}$\,s$^{-2}$ at solar neighbourhood after 2 Gyr which is negligible. The best measure of high mass satellites impact is the flaring of the disc although no strong heating is measured. Such insignificant heating suggests high mass subhaloes are not responsible for the observed heating. We conclude that 10-15\% contribution to the observed heating may originate from the impact of subhaloes. The next stage of this work is an exploration of the remaining processes which contribute towards the galactic disc heating, since we now have quantified the share from the infall of satellites using statistically relevant and high resolution analysis.      

\section*{Acknowledgements}
We thank Andrea Macci\`{o} and Volker Springel for providing us with VL\rom{2} and Aquarius simulation data. Also their fruitful comments and discussions were greatly useful. RM is funded by the German Research Foundation (DFG) under the Collaborative Research Center SFB 881 ``The Milky Way System'' through subproject A1/A2. Numerical simulations were performed on the Milky Way supercomputer, which is funded by the DFG SFB 881 subproject Z2 and hosted and co-funded by the J\"{u}lich Supercomputing Center (JSC).


\label{lastpage}
\end{document}